\documentclass{ametsocV6.1-nolinenumbers}

\def\p{\partial}
\def\f{\frac}

\def\ex{\bm{e_x}}

\def\ez{\bm{e_z}}

\def\vu{\bm{u}}
\def\tvu{\tilde{\bm{u}}}
\def\tTs{\tilde{T}_*}
\def\tT{\tilde{T}}
\def\tS{\tilde{S}}
\def\tP{\tilde{P}}
\def\tu{\tilde{u}}
\def\tv{\tilde{v}}
\def\tw{\tilde{w}}
\def\tm{\tilde{m}}
\def\tqT{\tilde{Q}_T}
\def\tqS{\tilde{Q}_S}
\def\tpt{\p_{\tilde{t}}}

\def\tpz{\p_{\tilde{z}}}
\def\tt{\tilde{t}}
\def\tx{\tilde{x}}
\def\ty{\tilde{y}}
\def\tz{\tilde{z}}
\def\tg{\tilde{\nabla}}
\def\tenu{\tilde{\nu}^e}
\def\tekT{\tilde{\kappa}_T^e}
\def\tekS{\tilde{\kappa}_S^e}
\def\tekTs{\tilde{\kappa}_*^e}
\def\ttnu{\tilde{\nu}^t}
\def\ttkT{\tilde{\kappa}_T^t}
\def\ttkS{\tilde{\kappa}_S^t}
\def\ttkTs{\tilde{\kappa}_*^t}

\title{Turbulent ice-ocean boundary layers in the well-mixed regime: \\ insights from direct numerical simulations}

\authors{Louis-Alexandre Couston\aff{a}\correspondingauthor{Louis-Alexandre Couston, louis.couston@ens-lyon.fr}}

\affiliation{\aff{a}{ENSL, UCBL, CNRS, Laboratoire de physique, F-69342 Lyon, France}}

\abstract{{The meltwater mixing line (MML) model provides a theoretical prediction of near-ice water mass properties that is useful to compare with observations.} If oceanographic measurements reported in a temperature-salinity diagram overlap with the MML prediction, then it is usually concluded that the local dynamics are dominated by the turbulent mixing of an ambient water mass with nearby melting ice. While the MML model is consistent with numerous observations, it is built on an assumption that is difficult to test with field measurements, especially near the ice boundary, namely that the effective (turbulent and molecular) salt and temperature diffusivities are equal. In this paper, this assumption is tested via direct numerical simulations of a canonical model for externally-forced ice-ocean boundary layers {in a uniform ambient}. {We focus on the well-mixed regime by considering an ambient temperature close to freezing and run the simulations until a statistical steady state is reached}. The results validate the assumption of equal effective diffusivities across most of the boundary layer. Importantly, the validity of the MML model implies a linear correlation between the mean salinity and temperature profiles normal to the interface that can be leveraged to construct a reduced ice-ocean boundary layer model based on a single scalar variable called thermal driving. We demonstrate that the bulk dynamics predicted by the reduced thermal driving model are in good agreement with the bulk dynamics predicted by the full temperature-salinity model. Then, we show how the results from the thermal driving model can be used to estimate the interfacial heat and salt fluxes{, and the melt rate}.\\ 
\begin{center}
\textit{This work has been submitted to the Journal of Physical Oceanography. Copyright in this work may be transferred without further notice.}
\end{center}}

\begin{document}

\maketitle

%
%
%
\statement{We investigate the turbulent dynamics and thermodynamical properties of water masses below ice shelves using new data from high-resolution simulations. This is important because observations of ice-ocean boundary layers are currently too scarce to construct reliable models of ice-shelf melting as functions of ocean conditions. Our results demonstrate that the turbulent diffusivities of salt and temperature are approximately equal in the well-mixed regime. This implies a linear correlation between the mean temperature and salinity profiles, consistent with the meltwater mixing line prediction that is often used to interpret polar observations. We take advantage of this correlation to propose a reduced model of ice-ocean boundary layers that can predict ice-shelf melt rates at relatively low computational cost.}
%
%
%

%

\section{Introduction}\label{sec1}

\begin{figure}[t]
  \centering
  \includegraphics[height=3.7cm]{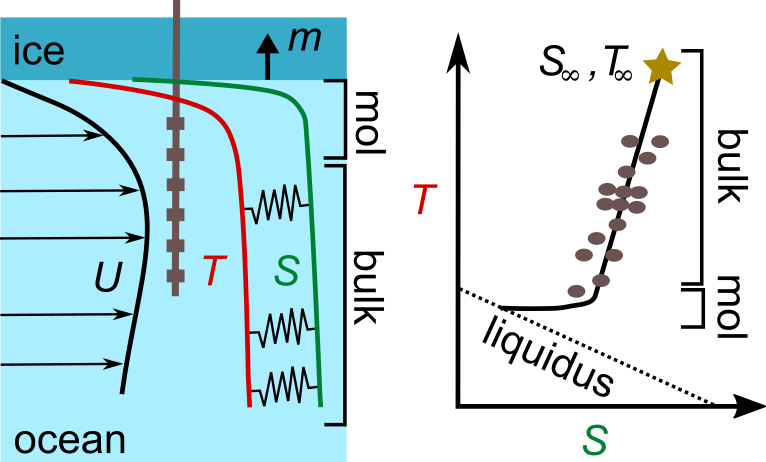}
  \put(-188,105){(a)}  
  \put(-85,105){(b)}  
  \caption{{Illustration of sub-ice water mass properties as functions of (a) depth and (b) in the $T-S$ diagram. mol and bulk highlight the molecular boundary layer and turbulent bulk, respectively. The springs in (a) illustrates the correlation between the mean $T$ and $S$ profiles predicted by Gade's MML theory and obtained when water mass transformations are dominated by the turbulent mixing of a uniform ambient $(T_{\infty},S_{\infty})$ with meltwater. The solid vertical gray line in (a) represents a conductivity, temperature and
depth (CTD) profiler, while the gray dots in (b) show an ensemble of plausible measurements. $m$ denotes the melt rate.}}\label{fig0}
\end{figure}

Melting and freezing processes transform water masses flowing next to ice shelves, icebergs and sea ice \citep{McDougall2014,Dinniman2016,Cenedese2023}.  {Ice melting produces meltwater that is typically colder and fresher than the ambient, such that the seawater temperature and salinity usually decrease toward the ice-ocean boundary. In the presence of a turbulent external flow, a well-mixed bulk emerges adjacent to a molecular diffusive boundary layer (cf. figure \ref{fig0}(a)). Multiple oceanographic observations have shown that the temperature $T$ and salinity $S$ are linearly correlated  (on average) in the turbulent bulk of an ice-ocean boundary layer (IOBL) flow \citep{Kimura2016,Stevens2020,Rosevear2022a,Davis2023}. In the $T-S$ diagram, this linear correlation becomes evident when the mean values of instantaneous measurements cluster on a straight line (figure \ref{fig0}(b)), which is then referred to as the meltwater mixing line (MML).}

{\cite{Gade1979} derived an analytical solution for the slope of the MML assuming steady state, a uniform ambient and a turbulent bulk. The slope, or correlation coefficient, relates the local temperature $T$ and salinity $S$ (in a time-averaged sense) to the constant temperature and salinity of the ambient $T_{\infty}$ and $S_{\infty}$ through
\begin{eqnarray}\label{eq:corr}
\mathcal{C} = \frac{T-T_{\infty}}{S-S_{\infty}},
\end{eqnarray}
and depends on parameters, such as latent heat of fusion, and ice properties, including ice salinity, whose variations in space and time are limited and most often negligible.
}


In the MML model \eqref{eq:corr}, the correlation is linear, i.e. $\mathcal{C}$ is a constant \citep[with value e.g. $\sim 2.7$ K/(g/kg) beaneath Pine Island Glacier ice shelf, see][]{Kimura2016}, for one key reason, which is that it assumes equal effective (turbulent and molecular) diffusivities for temperature and salt \citep{Gade1979}.\footnote{We note that the linear correlation also hinges on the subtle assumption that the ice melts in a vast amount of seawater, which is most likely appropriate for all oceanic applications, and refer to \cite{McDougall2014} for extensions of the theory to finite volumes of seawater.} The hypothesis that different scalars with different molecular diffusivities may have the same effective diffusivities has been tested in laboratory experiments \citep{Jackson2003,Martin2006} and numerical simulations \citep{Smyth2005,Ma2022b} replicating the open ocean environment. The aforementioned studies all show that the ratio of the effective temperature diffusivity to the effective salt diffusivity is indeed order unity when $Re_b \gg 100$, where $Re_b = \varepsilon/(\nu N^2)$ is the buoyancy Reynolds number ($\varepsilon$ is the viscous dissipation rate, $\nu$ is the kinematic viscosity and $N$ is the Brunt-Väisälä, or buoyancy, frequency), but that it increases (potentially above 2) with decreasing $Re_b \leq O(100)$ \citep{Gregg2018}. {In fact, at low $Re_b$, buoyancy becomes dynamically important \citep{Falor2023} and the turbulent Prandtl and Schmidt numbers depend on $Re_{\tau}$ but also $Pr$ and $Sc$, hence differ, such that diffential diffusion prevails \citep{Bouffard2013}. } 

{The MML model is not valid when turbulence is weak, spatially localized and/or intermittent, as in such cases molecular diffusivities influence the mean temperature and salinity profiles with depth. For instance, melting in a quiescient ambient leads to double-diffusive convection, with or without layering, and properties in a $T-S$ diagram that deviate from the MML prediction \citep{Kimura2015,Rosevear2022a}. The MML model is also insufficient to describe the properties of the meltwater mixture when the ambient is spatially stratified and/or changing over time, due to e.g. tidal flows or episodic subglacial discharge \citep{Davis2023}. In such cases, water mass properties span a two-dimensional subspace rather than a one-dimensional line in the $T-S$ diagram. Assessing the importance of melting relative to other phenomena in the water mass transformation process becomes limited, requiring, for instance, a composite tracer method for the calculation of the meltwater fraction \citep{Jenkins1999}.}

The linear correlation between $T$ and $S$, when it is valid, provides an opportunity to simplify the mathematical model governing the dynamics of {the IOBL}. The temperature can be inferred from the salinity and vice versa, such that it becomes unnecessary to solve the advection-diffusion equation for each scalar field. A natural (or practical) choice for the scalar field to be solved for in a reduced model of the IOBL is not $T$ nor $S$ but an aggregate variable called thermal driving, which is defined as \citep{Jenkins2011} 
\begin{eqnarray}\label{eq:tddef}
T_* = T-T_f(S,P_b) = T-(\lambda_1 S + \lambda_2 + \lambda_3 P_b). 
\end{eqnarray}
In equation \eqref{eq:tddef}, the freezing temperature $T_f$ is a function of salinity $S$ and reference pressure $P_b$ (chosen at the ice-ocean boundary), and is here linearized, as is the case in most oceanographic studies (with constants $\lambda_1$, $\lambda_2$, $\lambda_3$ listed in table \ref{tab:dimpar}). Thermal driving is the temperature above the local freezing point. Thus, at the ice-ocean interface it must satisfy $T_* = 0$ while in the far field $T_{*,\infty}=T_{\infty}-T_f(S_{\infty})$. Since temperature and salinity can be inferred from thermal driving using equations \eqref{eq:corr} and \eqref{eq:tddef}, a reduced thermal driving model, substituting the advection-diffusion equations for temperature and salt with a single advection-diffusion equation for $T_*$, can be used to reconstruct, or diagnose, the full IOBL properties in the turbulent bulk. The advantage of the reduced thermal driving model over the full $T-S$ model is significant. The diffusivity of thermal driving $\kappa_{*}$ can be taken much larger than the diffusivity of salt $\kappa_S$, such that the smallest-scale features of the reduced thermal driving model, given by the Batchelor scale $\ell_{B,*}$, become much coarser than the smallest-scale features of the full $T-S$ model, i.e., given by $\ell_{B,S}$ -- the ratio been given by $\ell_{B,*}/\ell_{B,S}\propto \sqrt{Le_*}$ with $Le_*=\kappa_{*}/\kappa_S$ \citep{Zhou2017}. Setting the molecular diffusivity of thermal driving equal to the diffusivity of temperature $\kappa_*=\kappa_T$, which is about 176 times larger than the diffusivity of salt \citep{Middleton2021}, leads to thermal driving {structures} roughly $\sqrt{176}\approx 13$ times coarser than salinity {structures}. A significant limitation of the thermal driving model is that $T$ and $S$ can be inferred from $T_*$ in the bulk only, i.e. not in the diffusive sublayers, since the linear correlation \eqref{eq:corr} only applies where turbulence dominates over molecular diffusion. A model for the diffusive sublayers of a turbulent subglacial ocean current thus becomes necessary to extrapolate the temperature and salinity profiles from the bulk to the ice-ocean interface \citep{Gade1979}. 

\cite{Jenkins2011} first introduced thermal driving as a useful prognostic variable in a Reynolds-averaged model of the IOBL{, i.e. wherein Reynolds stresses and fluxes are replaced by turbulent viscosity and diffusivities to focus on the smooth large-scale flow.} {The goal was to explore the effect of a stratified ambient. Thus, a second prognostic scalar field, density deficit in this case (relative to the ambient density), was kept in the model, which as a result had the same complexity as the original $T-S$ formulation \citep[see also][]{Hewitt2020}.} {A Reynolds-averaged thermal driving model with reduced complexity, i.e. with only one scalar equation assuming equal turbulent diffusivity, was subsequently proposed to study the transient evolution and stability of tilted IOBL under the effect of rotation, buoyancy, turbulent mixing and geostrophic currents \citep{Jenkins2016,Jenkins2021}.}

In this paper we achieve two goals. First, we use direct numerical simulations (DNS) of a model sub-ice ocean current to validate the hypothesis of equal effective diffusivities in the IOBL (\S\ref{sec2}). Second, we demonstrate that DNS results of a reduced thermal driving model can be used to reconstruct temperature and salinity fields consistent in the bulk with DNS results of a full $T-S$ model, and to estimate the heat and salt fluxes at the ice-ocean interface (\S\ref{sec3}). In so doing we provide evidence supporting the key hypothesis of the MML prediction and proof that the reduced thermal driving model is a worthwhile lightweight emulator of the IOBL dynamics, which we expect can expand the contribution of high-resolution simulations to studies of ice-ocean interactions \citep{Vreugdenhil2019,Mondal2019, Middleton2021,Rosevear2021,Begeman2022, Vreugdenhil2022,Rosevear2022b,Patmore2022}. We focus on the well-mixed regime for practical reasons \citep[also called shear-dominated regime, see][]{Rosevear2022b}. DNS are computationally expensive and the shear-dominated regime (weak density stratification) can be investigated in numerical domains much smaller than those required in the buoyancy-dominated regime (strong density stratification). {At relatively low $Re_{\tau}$, a shear-dominated flow is turbulent in both small and large channels (in the streamwise and spanwise directions), while a strongly-stratified flow is turbulent in large channels only, i.e. requiring a relatively broad numerical domain to avoid relaminarization, i.e. turbulence decay \citep{Garcia-Villalba2011}}. The MML prediction and thermal driving model are also expected to fail in the limit of strong density stratification, such that verifying their validity in the limit of weak stratification seems a useful first milestone.

%

\begin{table}[t]
\caption{List of the dimensional physical parameters relevant to this study. All values are standard and similar to the ones used in previous publications \citep{Jenkins2010,Vreugdenhil2019,Middleton2021}. They are appropriate for seawater below ice shelves up to several hundreds of meters thick.}\label{tab:dimpar}
\begin{center}
\begin{tabular}{lcc}
\hline\hline
Parameter name & Symbol & Value \\
\hline
  gravitational acceleration & $g$ & 9.81 m/s$^2$\\
  reference density & $\rho_0$ & 1000 g/kg \\
  seawater heat capacity & $c_p$ & 3974 J/K/kg \\
  latent heat of fusion & $L_i$ & $3.35\times 10^5$ J/kg \\
  salt-induced variation of the freezing temperature & $\lambda_1$ & $-5.73\times 10^{-2}$ K/(g/kg) \\
  surface freezing temperature of freshwater & $\lambda_2$ & $8.32\times 10^{-2}$ $^{\circ}$C \\
  pressure-induced variation of the freezing temperature & $\lambda_3$ & $-7.53\times 10^{-4}$ K/dbar \\
  kinematic viscosity & $\nu$ & $1.8\times 10^{-6}$ m$^2$/s \\
  thermal expansion coefficient & $\alpha$ & $3.87\times 10^{-5}$ 1/K \\
  haline contraction coefficient & $\beta$ & $7.86\times 10^{-4}$ 1/(g/kg) \\
  domain depth & $H$ & 1 m \\
  far-field salinity & $S_{\infty}$ & 35 g/kg \\
  far-field thermal driving & $T_{*,\infty}$ & $10^{-5}$ K \\
\hline
\end{tabular}
\end{center}
\end{table}




%

\section{Test of the equal effective diffusivities assumption}\label{sec2}

\subsection{Full $T-S$ model equations}

\begin{figure}[t]
  \noindent
  \includegraphics[height=5.7cm]{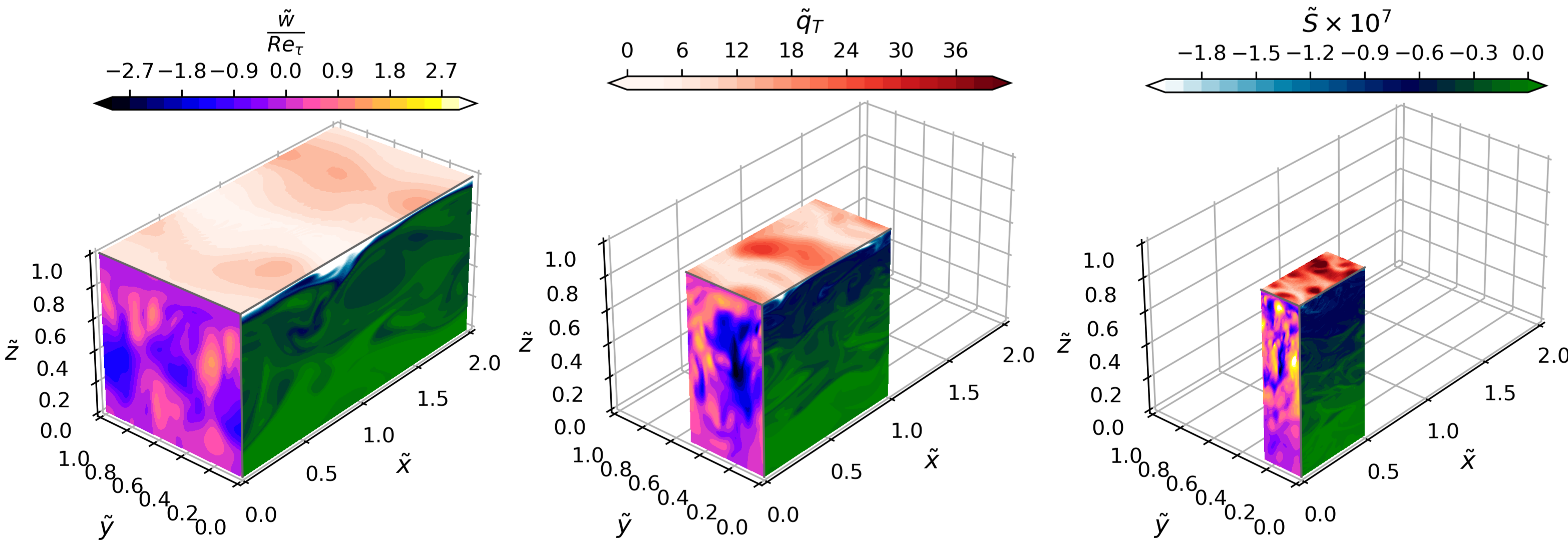}
  \put(-460,122){(a)}  
  \put(-310,125){(b)}  
  \put(-150,125){(c)}  
  \caption{Snapshots obtained at the final time of the simulations for $Pr=1$, $Le=10$ and (a) $Re_{\tau}=200$, (b) $Re_{\tau}=400$ and (c) $Re_{\tau}=800$. We show the normalized vertical velocity $\tw/Re_{\tau}$ on the upstream vertical slice ($\tx=0$), the normalized heat flux $\tilde{q}_T=[-\tpz \tT/(1-\langle \tT \rangle_{\perp})]|_{\tz=1}$ at the ice-ocean interface (a local version of $\tqT$ in \eqref{eq:fluxes}), and the salinity on the right side of the channel ($\ty=0$). The reduction of the domain size with $Re_{\tau}$ is obvious. The three colorbars apply to all snapshots.}\label{fig1}
\end{figure}

To test the equal effective diffusivities hypothesis we consider an externally-forced stratified sub-ice ocean flow in a horizontally periodic channel (cf. figure \ref{fig1}). We use a Cartesian coordinates system ($x,y,z$) with unit vector $\ez$ (bold variables denote vectors) upward opposite to gravity $\bm{g}=-g\ez$ to describe the dynamics and set the bottom of the channel as the origin of the $z$ axis. The ocean flow is forced by a uniform body force per unit volume $F$ (in kg.m$^{-2}$.s$^{-2}$) in the $\ex$ direction. We use the Boussinesq approximation. The reference density is $\rho_0$ and the density anomaly is a linear function of temperature and salinity \citep{Middleton2021}.  

{The  dimensional velocity vector, temperature, salinity and pressure are denoted by $\vu=(u,v,w)$, $T$, $S$ and $P$, respectively. For convenience, we solve the governing equations in dimensionless form. The dimensionless coordinates and variables, which we denote by tildes, are related to their dimensional counterparts through
\begin{eqnarray}\label{eq:scaling}
(\tilde{x},\tilde{y},\tilde{z})=\f{(x,y,z)}{H}, \quad \tilde{t}=\frac{t}{t_{\nu}}, \quad \tvu=\f{\vu}{u_{\nu}}, \quad \tT=\frac{T-T_f(S_{\infty},P_b)}{T_{*,\infty}}, \quad \tS = \f{S-S_{\infty}}{S_{\infty}}, \quad  \tP=\f{P-P_{st}}{p_{\nu}}, 
\end{eqnarray}
where $H$ is the channel thickness, $t_{\nu}=H^2/\nu$ is the viscous time scale, $u_{\nu}=\nu/H$ is the viscous velocity scale, $T_f(S_{\infty},P_b)$ is the freezing temperature (cf. equation \eqref{eq:tddef}) based on the far-field salinity ($S_{\infty}$) and constant hydrostatic pressure at the ice-ocean interface ($P_b$), $T_{*,\infty}=T_{\infty}-T_f(S_{\infty})$ is the far-field thermal driving, $P_{st}=P_b-\rho_0g(z-H)$ is the hydrostatic pressure and $p_{\nu}=\rho_0\nu^2/H^2$ is the dynamic pressure scale. Note that $\tS$ is a dimensionless salinity anomaly, which is always negative (since it is always lower than in the far field).}
%

{Substituting the dimensionless coordinates and variables \eqref{eq:scaling} into the IOBL bulk equations, which consist of the Navier-Stokes equations in the Boussinesq approximation and advection-diffusion equations for the temperature and salinity fields (resulting from the conservation of energy and salt), yields the dimensionless governing equations}
%
%
{
\begin{subequations}\label{eq:tsmodel}
\begin{eqnarray}\label{eq:tsmodel1}
&\tpt\tvu + \tvu\cdot\tg\tvu = -\tg\tP + \tg^2\tvu + Re_{\tau}^2 \ex - Re_{\tau}^2Ri_{\tau}(\tS-R_{\rho}\tT)\ez, \\ \label{eq:tsmodel2}
&\tg\cdot\tvu = 0, \\ \label{eq:tsmodel3}
&\tpt\tT + \tvu\cdot\tg\tT = Pr^{-1}\tg^2 \tT + \tilde{\mathcal{R}}_{T}, \\ \label{eq:tsmodel4}
&\tpt\tS + \tvu\cdot\tg\tS = Sc^{-1}\nabla^2 S + \tilde{\mathcal{R}}_{S}, 
\end{eqnarray}
\end{subequations}
}
with $Re_{\tau}$, $Ri_{\tau}$, $R_{\rho}$, $Pr$ and $Sc$ the dimensionless control parameters (see table \ref{tab:contpar} for the definition of all control parameters and the range of values considered). The rightmost terms in the temperature and salinity equations \eqref{eq:tsmodel3}-\eqref{eq:tsmodel4} are damping terms of the form $\tilde{\mathcal{R}}_{X}=Re_{\tau}^{3/2}\left(\f{\langle \tilde{X}\rangle_{\perp} - \tilde{X}_{\infty}}{2}\right)\left[1-\tanh\left(\f{\tz-0.2}{0.025}\right)\right]$, where $\langle \tilde{X} \rangle_{\perp}$ denotes the horizontal average and $\tilde{X}=\tT,\tS$. The damping terms force the mean temperature and salinity fields to the prescribed far-field values $\tT_{\infty}=1$ and $\tS_{\infty}=0$ where $\tz\lessapprox 0.2$. {The relaxation time scale is $Re_{\tau}^{-3/2}$ (in viscous time units), such that it is roughly inversely proportional to the mean ($Re_m$) and centreline ($Re_c$) Reynolds numbers (not shown; note that this is a steeper scaling than the classical $Re_m \sim Re_{\tau}^{8/7}$ scaling because of confinement effects, cf. \cite{Hwang2013}). Thus the dimensional relaxation time scale is inversely proportional to the far-field velocity, similar to previous studies \citep{Vreugdenhil2019}. Note that there is no damping of the velocity field.}

We limit our analysis to slowly melting ice such that the ice-ocean interface can be considered fixed at leading order (i.e. the simulation time is short relative to the melting time scale $\sim StPr$). The boundary conditions are free-slip (appropriate for a half channel flow), fixed temperature and fixed salinity on the bottom boundary, and no-slip, freezing temperature and melt-induced dilution at the top boundary, i.e., in dimensionless form 
%
%
\begin{subequations}\label{eq:tsbc}
\begin{eqnarray} \label{eq:tsbc2}
\tvu=(\tu,\tv,\tw)=\bm{0}, \quad \tT = \gamma \tS, \quad \tpz\tS = LeSt^{-1} \left( 1+\tS\right)\tpz\tT, \quad & \text{at} \quad \tz=1, \\ \label{eq:tsbc1}
\tpz\tu=\tpz\tv=\tw=0, \quad \tT = \tT_{\infty} = 1, \quad \tS = \tS_{\infty} = 0, \quad & \text{at} \quad \tz=0.  
\end{eqnarray}
\end{subequations}

{The melt-induced dilution equation (third equation in \eqref{eq:tsbc2}) arises from the combination of the energy and salt conservation equations at the interface, which, neglecting heat and salinity fluxes from the ice, can be written in dimensionless form as
\begin{eqnarray}\label{eq:ioicons}
\tm St=-Pr\tpz\tT, \quad
\tm\left( 1+\tS\right)=-Sc\tpz\tS, \quad & \text{at} \quad \tz=1, 
\end{eqnarray}
with $\tm$ the dimensionless melt rate}. Note that we neglect meltwater advection due to the density difference between liquid water and solid ice when we write the no-slip condition \eqref{eq:tsbc2} at $\tz=1$. This approximation  is valid in the limit of infinitesimal thermal driving, which we focus on, but can lead to significant errors when thermal driving is of order 1$^{\circ}$C or more as it neglects the vertical transport of sensible heat due to the melt-induced up-welling \citep{Jenkins2001}. {This transport term is negligible in our simulations because it is approximately equal to $\tilde{m}\tilde{T}$, which is of order $10^{-7}$ or smaller in all cases, hence negligible compared to the turbulent heat flux that is of order $O(10)$.}   

\begin{table}[t]
\caption{List of the dimensionless control parameters appearing in the governing equations \eqref{eq:tsmodel}-\eqref{eq:tsbc}. We provide typical values for the ocean besides those considered in this study. The list of physical parameters used to construct the dimensionless parameters can be found in table \ref{tab:dimpar}. The ocean values consider a friction velocity between 0 and 0.18 m/s and thermal driving between 0 and 10 K {(other parameters taken from table \ref{tab:dimpar})}. Note that the dimensionless domain lengths $L_x$ and $L_y$ are two other control parameters that are changed across simulations (see table \ref{tab:numexp}). {The friction Richardson number is based on a salinity gradient as salinity differences typically dominate density differences.}}\label{tab:contpar} 
\begin{center}
\begin{tabular}{lccccl}
\hline\hline
Parameter name & Symbol & Definition & Ocean value(s) & Study value(s) & Comments \\
\hline
  friction Reynolds & $Re_{\tau}$ & $\sqrt{\frac{F H^3}{\rho_0\nu^2}}$ & $0$ to $10^5$  & 200 to 800 & \\
  friction Richardson & $Ri_{\tau}$ & $\frac{g\rho_0\beta S_{\infty}}{F}$ & 0 to $+\infty$ & $1.3\times10^5$ to $2.1\times10^6$ & Salt Rayleigh number $Ra_S=Re_{\tau}^2Ri_{\tau}Sc$ \\
  density ratio & $R_{\rho}$ & $\frac{\alpha T_{*,\infty}}{\beta S_{\infty}}$ & 0 to $1.4\times10^{-2}$ & $1.4\times10^{-8}$ &  \\
  Prandtl & $Pr$ & $\frac{\nu}{\kappa_T}$ & $\approx 14$ & 1 to 10 \\
  Schmidt & $Sc$ & $\frac{\nu}{\kappa_S}$ & $\approx 2500$ & 1 to 100 & Lewis number $Le=Sc/Pr$ \\
  salt-induced freezing slope & $\gamma$ & $\frac{\lambda_1S_{\infty}}{T_{*,\infty}}$ & $-\infty$ to $-0.2$ & $-2.0\times10^5$ &  \\
  Stefan & $St$ & $\frac{L_i}{c_wT_{*,\infty}}$ & $8.4$ to $+\infty$ & $8.4\times10^6$ \\
\hline
\end{tabular}
\end{center}
\end{table}



\subsection{Numerical experiment}

The governing equations \eqref{eq:tsmodel}-\eqref{eq:tsbc} of the full $T-S$ model are solved numerically using the open-source pseudo-spectral code Dedalus \citep{Burns2020}. We use Fourier mode decompositions in the horizontal directions and Chebyshev mode decomposition in the vertical direction. We use a third-order multi-stage explicit-implicit Runge-Kutta scheme for time integration and we set the Courant-Friedrich-Lewy parameter to 0.5. We run {11} simulations with varying $Re_{\tau}$, $Pr$ and $Le$ in order to test the equal effective diffusivities hypothesis underlying the MML theory. Owing to computational limitations, {we cannot run simulations with $1\ll Pr \ll Le$, which is the parameter range relevant to the natural ocean.} In particular, the Schmidt number (or inverse of the normalized salt diffusivity) does not exceed 100, hence is always less than the ocean value $Sc\approx 2464$ \citep{Middleton2021}. However, the MML theory does not depend on any particular value of $Sc$ (or $Pr$ and $Le$) and we will show that the equal effective diffusivity hypothesis is in fact valid for all $(Re_{\tau},Pr,Le)$ tested{, including cases with $Pr=10$ and $Le=10$ that have the same hierarchy of scales as the natural ocean (i.e. large momentum structures, medium-size thermal structures and small salinity structures).} {The resolution of the simulations is discussed in the Supplementary Material. In brief, the simulations adequately resolve the diffusive sublayers, thanks to the clustering of Chebyshev points near boundaries, as well as the Kolmogorov length scale. The Batchelor length scale is not always well resolved, especially at high $Le$. However, running higher-resolution simulations has shown that the flow statistics are not sensitive to the relatively coarse resolution of the simulations in the bulk.}

We focus on the weakly-stratified limit of the IOBL. We ensure that the simulations remain turbulent at all times (no relaminarization) by consideing a very small far-field thermal driving $T_{*,\infty}=10^{-5}$ K (and a far-field salinity $S_{\infty}=35$ g/kg). All simulations are first run without the scalar fields until a statistical steady state is reached. Then the governing equations for the scalar fields are solved for and coupled to the hydrodynamics. The simulations are run for a relatively long time (up to 100 friction time units) to guarantee that a statistical steady-state is reached and enough data is accumulated to compute reasonably well converged statistics. Table \ref{tab:numexp} lists all simulation cases and parameters.

For computational expediency, we run the simulations in a minimal channel flow unit (MCU), i.e., considering short and narrow channels \citep{Jimenez1991,Flores2010}. We hypothesize that the validity of the equal turbulent diffusivities hypothesis in an MCU implies the validity in an unconfined channel. This hypothesis is supported by the fact that the small-scale turbulence remains healthy in an MCU, especially near the walls, despite well-known horizontal confinement effects in the outer layer \citep{Flores2010,Hwang2013,LozanoDuran2014,DeGiovanetti2016}. It is also supported by the fact that one test case {run} in both an MCU and a larger domain (two times larger in both horizontal directions) resulted in equal turbulent diffusivities irrespective of the domain size. {The difference between the drag coefficient $C_D$ (defined in the next section) obtained for a MCU and a domain twice as large when $Re_{\tau}=400$, $Pr=1$ and $Le=10$, which we report in the last column of the eighth and last lines of table \ref{tab:numexp}, highlights the impact that the (small) horizontal lengths have on the global dynamics (but not on the equal turbulent diffusivity hypothesis), in agreement with previous studies \citep{Hwang2013,DeGiovanetti2016}.}  

\subsection{Diagnostics}

{Our analysis of the simulations focuses on horizontal-plane and volume averages of the variables and on vertical momentum, temperature and salinity fluxes. We write $\langle \cdot \rangle_{\perp}$, $\langle \cdot \rangle_{\tz}$ and $\langle \cdot \rangle$ the horizontal, vertical and volume averaging operators, respectively, and $\overline{ \;\cdot\; }$ the time averaging operator (computed at statistical steady state).}

{We define the dimensionless horizontally-averaged heat and salt fluxes as
%
%
\begin{eqnarray}\label{eq:fluxes}
\tqT=\langle \tw\tT-Pr^{-1}\tpz\tT\rangle_{\perp}, \quad \tqS=\langle \tw\tS-Sc^{-1}\tpz\tS\rangle_{\perp},
\end{eqnarray}
and note that the convective fluxes in \eqref{eq:fluxes} are zero at the no-slip top and impermeable bottom boundaries. The vertical transport efficiency of momentum, heat and salinity is evaluated from the temperature Nusselt number, salinity Nusselt number and drag coefficient, defined as
%
%
\begin{eqnarray}\label{eq:output}
Nu_T = \frac{\overline{ \tqT(\tz=1) }}{\tqT^{diff}}, \quad Nu_S = \frac{\overline{ \tqS(\tz=1) }}{\tqS^{diff}}, \quad C_D = 2\left(\frac{Re_{\tau}}{\langle\overline{ \tu }\rangle}\right)^2,
\end{eqnarray}
%
where 
\begin{eqnarray}\label{eq:difffluxes}
\tqT^{diff}=-Pr^{-1}\left(\langle\overline{ \tT(\tz=1) }\rangle_{\perp}-1\right), \quad \tqS^{diff}=-Sc^{-1}\langle\overline{ \tS(\tz=1) }\rangle_{\perp},
\end{eqnarray}
are the heat and salt fluxes of the equivalent purely-diffusive state ($\tvu=\tilde{\bm{0}}$). We recall that $\tT(\tz=0) = 1$ and $\tS(\tz=0) = 0$ by definition and that the channel height is unity.  
} 

{
We test the validity of the meltwater mixing line theory by computing the effective Prandtl and Lewis numbers (superscripts $e$), defined as
\begin{eqnarray}\label{eq:lewis}
Pr^e = \f{\tenu}{\tekT}, \quad Le^e = \f{\tekT}{\tekS},
\end{eqnarray}
with $\tenu$, $\tekT$, $\tekS$ (resp. $\ttnu$, $\ttkT$, $\ttkS$) the effective (resp. turbulent; superscripts $t$) viscosity and temperature and salt diffusivities, which we write as
\begin{eqnarray}\label{eq:turbdiff}
& \tenu = \ttnu + 1 = \f{\langle\overline{ \tw\tu - \tpz\tu}\rangle_{\perp}}{-\langle\overline{ \tpz\tu }\rangle_{\perp}}, \quad \tekT = \ttkT + Pr^{-1} = \f{\overline{ \tqT }}{-\langle\overline{ \tpz\tT }\rangle_{\perp}}, \quad \tekS = \ttkS + Sc^{-1} = \f{\overline{ \tqS }}{-\langle\overline{ \tpz\tS }\rangle_{\perp}}. 
\end{eqnarray}
The convective terms in \eqref{eq:turbdiff} involve the \textit{full} variables; however, they could be replaced by their fluctuating counterparts since $\langle\overline{\tw}\rangle_{\perp}=0$ $\forall \tz$ at statistical steady state. Finally, the buoyancy Reynolds number reported in table \ref{tab:numexp} is defined as 
\begin{eqnarray}\label{eq:reb}
Re_b = \langle\frac{\langle\overline{\tilde{\epsilon}}\rangle_{\perp}Sc}{Ra_S\langle\overline{-\tpz\tS+R_{\rho}\tpz\tT}\rangle_{\perp}}\rangle_{\tz >0.5},
\end{eqnarray}
with $\langle\cdot\rangle_{\tz>0.5}$ denoting the average over the upper half of the domain and $\langle\overline{\tilde{\epsilon}}\rangle_{\perp}$ the mean (pseudo) turbulent kinetic energy dissipation rate (cf. Supplementary Material). Here we find that the buoyancy Reynolds number is always much larger than O(100) (cf. table \ref{tab:numexp}), such that we might expect equal turbulent diffusivities in all our simulations.
}

\begin{table}[t]
\caption{{List of key input parameters (physical and numerical) and output variables for all simulation cases. $(n_x,n_y,n_z)$ is the number of Fourier and Chebyshev modes in all three directions (not including the scaling factor 3/2 used for dealiasing), $Le^{e}$ is the effective Lewis number based on the effective diffusivities of temperature and salt in the bulk, i.e., averaged between $\tz=0.3$ and $\tz=0.7$ and $Re_b$ is the buoyancy Reynolds number given by equation \eqref{eq:reb}. For the reduced $T_*$ simulations $Nu_T$ is the Nusselt number with $\tTs$ replacing $\tT$. The far-field thermal driving and salinity are set to $T_{*,\infty}=10^{-5}$ K and $S_{\infty}=35$ g/kg in all cases. Simulation name \textit{Full wide} on the last line refers to a \textit{Full} $T-S$ simulation with double the length in the horizontal directions.}}\label{tab:numexp}
\begin{center}
\begin{tabular}{cccccccccccc}
\hline\hline
Type & $Re_{\tau}$ & $Pr$ ($Pr_*$) & $Le$ & $(L_x,L_y,L_z)$ & $(n_x,n_y,n_z)$ & $Le^{eff}$ & $Re_b$ & $Nu_T$ & $Nu_S$ & $C_D$ \\
\hline
Full $T-S$ & 200 & 1 & 1 & (2,1,1) & (64,64,48)  & 1.000 & $10^4$ & 8.6 & 8.6 & 0.0080 \\
Full $T-S$ & 200 & 1 & 10 & (2,1,1) & (128,128,96)  & 0.997 & $10^4$ & 8.4 & 27.3 & 0.0078 \\
Full $T-S$ & 200 & 1 & 30 & (2,1,1) & (256,256,192) & 1.002 & $10^4$ & 8.1 & 41.5 & 0.0079 \\
Full $T-S$ & 200 & 1 & 100 & (2,1,1) & (256,256,192) & 1.002 & $2\times 10^4$ & 8.3 & 63.7 & 0.0079 \\
{Full $T-S$} & 200 & 5 & 5 & (2,1,1) & (256,256,192) & 0.991 & $2\times 10^4$ & 27.2 & 65.0 & 0.0080 \\
Full $T-S$ & 200 & 10 & 10 & (2,1,1) & (256,256,192) & 0.992 & $4\times 10^4$ & 20.1 & 39.9 & 0.0080 \\
Full $T-S$ & 400 & 1 & 1 & (1,0.5,1) & (64,64,96)  & 1.000 & $10^5$ & 13.9 & 13.9 & 0.0056 \\
Full $T-S$ & 400 & 1 & 10 & (1,0.5,1) & (128,128,192) & 0.996 & $10^5$ & 14.3 & 54.1 & 0.0057 \\
Full $T-S$ & 400 & 10 & 10 & (1,0.5,1) & (256,256,384) & 1.000 & $3\times 10^5$  & 54.2 & 137.3 & 0.0056 \\
Full $T-S$ & 800 & 1 & 1 & (0.5,0.25,1) & (64,64,128) & 1.000 & $2\times 10^6$  & 19.7 & 19.7 & 0.0031 \\
Full $T-S$ & 800 & 1 & 10 & (0.5,0.25,1) & (128,128,256) & 1.011 & $2\times 10^5$ & 20.2 & 95.0 & 0.0031 \\
Reduced $T_*$ & 200 & 1 & - & (2,1,1) & (64,64,48) & -  & - & 8.7 & - & 0.0078 \\
{Reduced $T_*$} & 200 & 5 & - & (2,1,1) & (128,128,96) & - & -  & 20.1 & - & 0.0079 \\
Reduced $T_*$ & 200 & 10 & - & (2,1,1) & (128,128,96) & - & -  & 28 & - & 0.0080 \\
Reduced $T_*$ & 400 & 1 & - & (1,0.5,1) & (64,64,96) & - & -  & 14.0 & - & 0.0056 \\
Reduced $T_*$ & 400 & 10 & - & (1,0.5,1) & (128,128,192) & - & -  & 53 & - & 0.0056 \\
Reduced $T_*$ & 800 & 1 & - & (0.5,0.25,1) & (64,64,128) & - & -  & 19.5 & - & 0.0032 \\
Full wide & 400 & 1 & 10 & (2,1,1) & (256,256,192) & 0.985 & $10^5$ & 16.3 & 55.1 & 0.0062 \\
\hline
\end{tabular}
\end{center}
\end{table}

\subsection{Results}

\begin{figure}[t]
  \noindent
  \includegraphics[height=5.4cm]{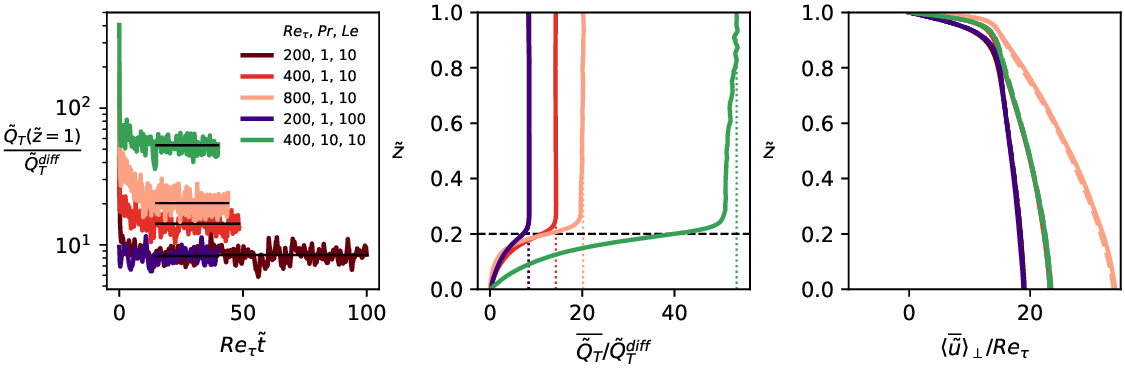}
  \put(-435,158){(a)}    
  \put(-285,158){(b)}    
  \put(-135,158){(c)}  
  \caption{(a) Normalized heat flux at the ice-ocean interface as a function of time (in friction time units) for five cases. The horizontal black lines show the temporal windows used for time averaging. (b) Normalized time-averaged heat flux as a function of depth. The vertical dotted lines highlight the normalized heat flux at the ice-ocean interface ($\tz=1$) while the horizontal dashed line highlights the upper limit of the damping region. Note that the $(200,1,100)$ and $(200,1,10)$ results overlap. (c) Normalized time-averaged horizontal velocity as a function of depth. The velocity profiles with the same $Re_{\tau}$ overlap. Results from the reduced $T_*$ model are also shown (dashed lines) but are indistinguishable from the full $T-S$ model results (solid lines).}\label{fig2}
\end{figure}

To illustrate the relatively long time of the simulations, we first show in figure \ref{fig2}(a) the time series of the normalized and horizontally-averaged heat flux at the ice-ocean interface $\tqT (\tz=1)/\tqT^{diff}$. We show the results from five simulations only for clarity. The temporal windows used for time averaging are highlighted by the horizontal black lines and last at least 25 friction time units. The time-averaged normalized heat flux is then shown as a function of $\tz$ in figure \ref{fig2}(b). The heat flux is almost depth invariant for $\tz\geq 0.2${, in agreement with the time-averaged heat equation at statistical steady state (without the relaxation term).} For the most ambitious simulation ($Re_{\tau},Pr,Le$)=(400,10,10) we observe a $\sim$10\% deviation with depth {(suggesting that the boundary layer is not perfectly equilibrated yet),} which we expect would decrease with increasing simulation time and statistical averaging. For $\tz\leq 0.2$ the source/sink of heat comes from the relaxation term $\tilde{\mathcal{R}}_T$, which is not included in the definition of ${\tqT}$. As expected, $\overline{\tqT}\rightarrow 0$ as $\tz\rightarrow 0$ because of the no penetration condition at $\tz=0$ and the imposed uniform mean temperature profile in the damping region. {Figure \ref{fig2}(c) shows the mean horizontal velocity as a function of depth. As expected, we observe a boundary layer close to the (top) ice-ocean interface and a maximum at $\tilde{z}=0$ where the flow is shear free. The dashed lines show the mean horizontal velocity obtained in reduced $T_*$ simulations (described in \S\ref{sec3}). They overlap perfectly with the solid lines, which show the full $T-S$ simulations results. This overlap does not yet prove that the reduced thermal-driving model can replace the full $T-S$ model as we are running the simulations in the shear-dominated regime, such that the scalar fields are almost passive and have little influence on the dynamics. The decrease of the drag coefficient with $Re_{\tau}$ (reported in table \ref{tab:numexp}) is consistent with the increase of the normalized velocity in figure \ref{fig2}(c). It is partly due to confinement effects, the higher $Re_{\tau}$ simulations having smaller domain sizes \citep[see figure 4 in][]{Hwang2013}.}  

\begin{figure}[t]
  \noindent
  \includegraphics[height=5.4cm]{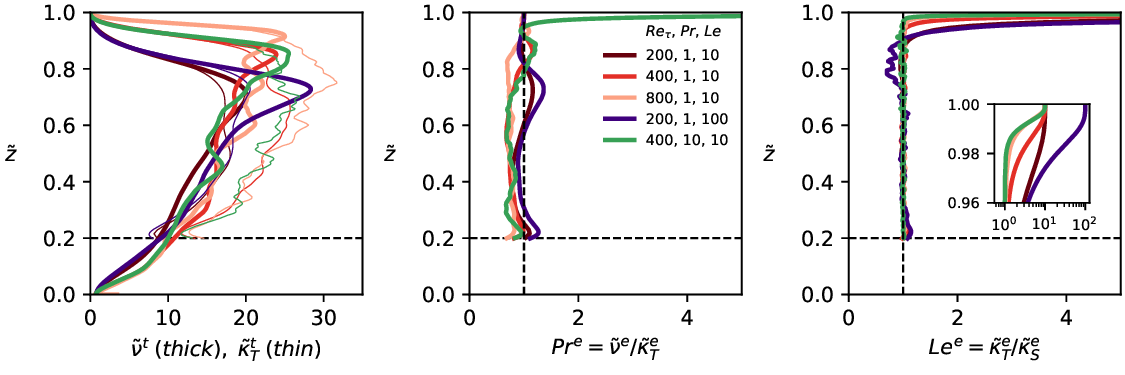}
  \put(-445,158){(a)}  
  \put(-290,158){(b)}  
  \put(-135,158){(c)}  
  \caption{(a) Turbulent viscosity (thick solid lines) and temperature diffusivity (thin) as a function of depth, (b) effective Prandtl number as a function of depth, and (c) effective Lewis number as a function of depth. We only show five cases for clarity. The vertical dashed lines highlight ratios equal to 1. Note that the turbulent temperature diffusivity is not shown in the relaxation region $\tz\leq 0.2$ (below the horizontal dashed lines) because the damping terms generate depth-invariant temperature (and salinity) profiles with unrealistically large turbulent diffusivities. By construction, $Pr^{e}\rightarrow Pr$ and $Le^{e}\rightarrow Le$ when $\tz\rightarrow 1$ {(see the inset in (c) zooming on the top diffusive sublayer).}}\label{fig3}
\end{figure}

We show the vertical profiles of $\ttnu$, $\ttkT$ and $\ttkS$ in figure \ref{fig3}(a) (masking the damping region $\tz\leq 0.2$ for clarity). In all cases, $\ttnu \gg 1$, $\ttkT \gg 1$ and $\ttkS \gg 1$ when $0.2 \leq \tz\leq 0.9$, demonstrating the turbulent nature of the flow and predominance of convective transports over molecular diffusion in the bulk. The turbulent viscosity reaches a maximum relatively close to the ice-ocean interface, where the shear is large, and then decreases with depth, in agreement with previous studies. The turbulent diffusivities reach similarly large values close to the top boundary but then decay more slowly with depth. {Consequently, $Pr^e$ and $Sc^e$ become most often less than one in the bulk (as observed in figure \ref{fig3}(b) for $Pr^e$), in agreement with previous models and observations of turbulent Prandtl and Schmidt numbers when $Pr,Sc\geq 1$ \citep{Reynolds1975,Alcantara-Avila2021}}. The key result of this section is that the effective Lewis number $Le^{e}=\tekT/\tekS$ is close to one in most of the bulk in all cases (see figure \ref{fig3}(c)).  

\begin{figure}[t]
  \centering
  \includegraphics[height=5.4cm]{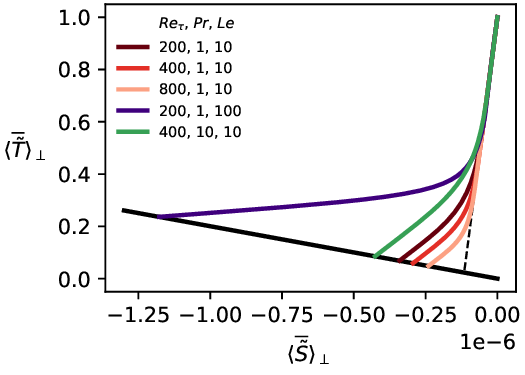}
  \caption{Mean temperature shown as a function of mean salinity (both are depth-dependent variables) for five cases. The thick solid black line shows the freezing temperature while the dashed black line shows the meltwater mixing line prediction.}\label{fig4}
\end{figure}

To further demonstrate the validity of the MML theory in the weakly-stratified limit, we show the mean temperature profile $\langle \overline{ \tT } \rangle_{\perp}$ as a function of the mean salinity profile $\langle \overline{ \tS } \rangle_{\perp}$ in figure \ref{fig4} (the $T-S$ diagram). All simulation results perfectly overlap with the MML prediction in the bulk.\footnote{Note that this is true also near the bottom boundary (top of the curves) because of the imposed relaxation terms. Without relaxation terms the uniform far-field is replaced by a boundary layer with $T-S$ correlations that are different from the MML prediction, unless the bottom boundary is also an ice-ocean interface and buoyancy effects are negligible.} The correlation between $\langle \overline{\tT} \rangle_{\perp}$ and $\langle \overline{\tS} \rangle_{\perp}$ is always the same far from the ice-ocean interface, i.e. the curves all have the same slope, regardless of the control parameters $Re_{\tau}$, $Pr$ and $Le$, in agreement with the MML theory (derived in appendix).

\begin{figure}[t]
  \centering
  \includegraphics[height=5.4cm]{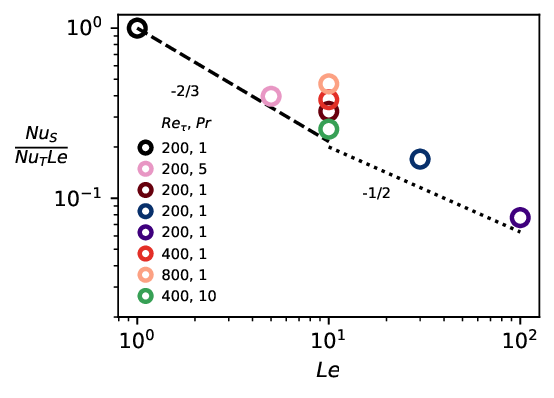}
  \caption{{Nusselt number ratio shown as a function of $Le$ and normalized by $Le$ such that it corresponds to the ratio of salt to temperature transfer velocities in \cite{Holland1999}.}}\label{fig4b}
\end{figure}

{The $\langle \overline{\tT} \rangle_{\perp}$ profiles as functions of $\langle \overline{\tS} \rangle_{\perp}$ differ between cases as they approach the freezing line. The salinity deficit at the interface, i.e. $-\overline{\tilde{S}}(\tilde{z}=1)$, increases with $Le$ (compare $Le=10$ and $Le=100$ for $Pr=1$ and $Re_{\tau}=200$) and $Pr$ (compare $Pr=1$ and $Pr=10$ for $Le=10$ and $Re_{\tau}=400$) but decreases with $Re_{\tau}$ (dark to light red curves). This parametric dependence is consistent with previous studies, once we realize that the interfacial salinity is inversely proportional to the ratio of Nusselt numbers, and thus to the ratio of the salt to heat exchanges velocities, denoted by $\gamma_S$ and $\gamma_T$ (m.s$^{-1}$), which are parameterized in ISOBL models \citep{Holland1999}. In fact, $\gamma_{T} = \f{Nu_{T}\kappa_{T}}{H}$ and $\gamma_{S} = \f{Nu_{S}\kappa_{S}}{H}$ by definition, such that 
\begin{eqnarray}\label{eq:evratio}
\frac{\gamma_S}{\gamma_T}=\frac{Nu_S}{Nu_TLe}=\frac{\overline{\tilde{Q}_T}(\tz=1)\left[ 1-\overline{\tilde{T}}(\tz=1) \right]}{\overline{\tilde{Q}_S}(\tz=1)\left[-\overline{\tilde{S}}(\tz=1)\right]}=\frac{St\left[ 1-\overline{\tilde{T}}(\tz=1) \right]}{-\left[1+\overline{\tilde{S}}(\tz=1)\right]\overline{\tilde{S}}(\tz=1)}\approx \frac{St}{-\overline{\tilde{S}}(\tz=1)}.
\end{eqnarray}
The second to last expression in \eqref{eq:evratio} combines equations \eqref{eq:ioicons} and \eqref{eq:fluxes} while the last expression is obtained in the limit $\overline{\tilde{S}}(\tilde{z}=1) \rightarrow 0$. We show $Nu_S/(Nu_TLe)$, i.e. the exchange velocity ratio, as a function of $Le$ in figure \ref{fig4b}. We clearly observe a decrease of $Nu_S/(Nu_TLe)$ with $Le$ and a (relatively) more modest increase with $Re_{\tau}$ for $Le=10$ and $Pr=1$, consistent with the $Le$- and $Re_{\tau}$-dependence of the interfacial salinity. The $Le$ dependence of $Nu_S/(Nu_TLe)$ can be modelled with a power law with a -2/3 to -1/2 exponent, in good agreement with the parameterizations for $\gamma_S$ and $\gamma_T$ used in \cite{Holland1999} for shear-driven flows. Starting from equations (11)-(12) in \cite{Holland1999} we find that $\gamma_S/\gamma_T$ scales like $ Le^{-2/3}$ in the large $Pr$, large $Sc$ and (relatively) small $Re_{\tau}$ limit. The $Le$ dependence of $Nu_S/(Nu_TLe)$ is also consistent with the -2/3 (resp. -1/2) power law scaling uncovered in recent simulations of multicomponent vertical convection at relatively low $Le$ (resp. high $Le$) \citep{Howland2023}. The decrease of $-\overline{\tilde{S}}(\tilde{z}=1)$ with $Re_{\tau}$ and increase with $Pr$ observed in figure \ref{fig4} are also in good agreement with the exchange velocity formula in \cite{Holland1999} since $\gamma_S$ (resp. $\gamma_T$) increase with $Re_{\tau}$ and $Pr$ (resp. $Sc$) such that the ratio $\gamma_S/\gamma_T$ decreases with $Pr$ (as long as $Le > 1$) but increases with $Re_{\tau}$.}

\section{Test of the reduced thermal driving model}\label{sec3}

\subsection{Mathematical derivation}

In this section we introduce the governing equations for the reduced thermal driving model and the formula to reconstruct the temperature and salinity fields in the bulk using dimensionless variables. We substitute the temperature $\tT$ and salinity $\tS$ with a single scalar field, called thermal driving, which we define as $\tTs=\tT-\gamma \tS$, i.e. as the temperature above the local freezing point. From the MML model, which predicts $\tT-1=\tS St$ (see appendix) and the definition of $\tTs$, we can rewrite the buoyancy term $(\tS-R_{\rho}\tT)$ as $\tTs\left(\frac{1-R_{\rho}St}{St-\gamma}\right)+\frac{\gamma R_{\rho}-1}{St-\gamma}$ and discard the last term since it is constant {($\alpha$ and $\beta$ are constants, see table \ref{tab:dimpar})} and can be balanced by the hydrostatic pressure. The governing equations for the reduced thermal driving model become
%
\begin{subequations}\label{eq:tdmodel}
\begin{eqnarray}
&\tpt\tvu + \tvu\cdot\tg\tvu = -\tg\tP + \tg^2\tvu + Re_{\tau}^2 \vec{e_x} - Re_{\tau}^2Ri_{\tau}\tTs\left(\frac{1-R_{\rho}St}{St-\gamma}\right)\vec{e_z}, \\
&\tg\cdot\tvu = 0, \\
&\tpt\tTs + \tvu\cdot\tg\tTs = 
\frac{1}{Pr_*}\tg^2\tTs + \tilde{\mathcal{R}}_{T_*}, 
\end{eqnarray}
\end{subequations}
with $Pr_*$ the Prandtl number for the scalar field $\tTs$ and $\tilde{\mathcal{R}}_{T_*}$ a relaxation term similar to  $\tilde{\mathcal{R}}_{T}$ in equation \eqref{eq:tsmodel3}. The kinematic boundary conditions are the same as for the full $T-S$ model, and the boundary conditions for thermal driving are simply
\begin{eqnarray}
& \tTs(\tz=1) = 0, \quad \tTs(\tz=0) = \tT_{*,\infty} = 1. 
\end{eqnarray}

\subsection{Numerical experiment}

We solve equations \eqref{eq:tdmodel} (together with the boundary conditions) numerically with a pseudo-spectral solver (again from the Dedalus code) similar to the one used to solve equations \eqref{eq:tsmodel}-\eqref{eq:tsbc} of the full $T-S$ model (similar spectral expansions and time stepper). We run simulations of the reduced thermal driving model for $Re_{\tau}=200,$ 400, 800 and $Pr_*=1$, 5, 10 (see the list of simulation cases in table \ref{tab:numexp}) and compare the results with the output of the simulations of the full $T-S$ model (with similar control parameters). The final checkpoint file of each thermal driving simulation was used to initialize a new $T-S$ simulation (splitting the thermal driving field into temperature and salinity fields according to equations \eqref{eq:bulkrec}) with $Le=1$, 10 or 100 -- i.e. distinct from the original $T-S$ simulations. This procedure was designed to test the sensitivity of the $T-S$ simulations to the initial conditions for temperature and salinity (uniform or from thermal driving at statistical steady state). However, we did not observe any significant sensitivity to initial conditions, such that we do not include these \textit{split} simulations in table \ref{tab:numexp}.

\subsection{Diagnostics}

{We assess the reliability of the reduced $T_*$ model by comparing the mean temperature and salinity profiles, and the mean interfacial fluxes, diagnosed from $\tilde{T}_*$ results, with those obtained in full $T-S$ simulations. The reconstruction of the temperature and salinity from $\tilde{T}_*$ combines the definition of thermal driving and the MML (see details in appendix) and yields for the diagnosed fields (denoted by superscript \textit{diag})
\begin{eqnarray}\label{eq:bulkrec}
 \tT^{diag} = \frac{St \langle \overline{ \tTs } \rangle_{\perp}-\gamma}{St-\gamma}, \quad \tS^{diag} = \frac{\langle \overline{ \tTs } \rangle_{\perp}-1}{St-\gamma}.
\end{eqnarray}
The interfacial heat and salt fluxes are then diagnosed provided that we know the effective (or turbulent) diffusivity $\tekTs$ ($\ttkTs$) of the $\tilde{T}_*$ field, defined as
\begin{eqnarray}\label{eq:turbtddiff}
\tekTs  = \ttkTs + Pr_*^{-1} = \frac{\langle \overline{\tw\tTs-Pr_*^{-1}\tpz\tTs}\rangle_{\perp}}{\langle -\overline{\tpz\tTs} \rangle_{\perp}}.
\end{eqnarray}
If all scalar effective diffusivities are equal, i.e. $\tekT \approx \tekS \approx \tekTs$ (in the bulk), and the fluxes are depth invariant, we can indeed estimate for a uniform ambient and at statistical steady state 
\begin{subequations}\label{eq:fluxrec}
\begin{eqnarray}
& \tqT^{diag}(\tz=1) = -\tekT\langle \overline{\tpz\tT^{diag}} \rangle_{bulk} = -\frac{\tekTs St}{St-\gamma} \langle \overline{\tpz\tTs} \rangle_{bulk}, \\
& \tqS^{diag}(\tz=1) = -\tekS\langle \overline{\tpz\tS^{diag}} \rangle_{bulk} = -\frac{\tekTs}{St-\gamma} \langle \overline{\tpz\tTs} \rangle_{bulk},
\end{eqnarray}
\end{subequations}
where $\langle\cdot \rangle_{bulk} =\f{1}{0.7-0.3}\int_{0.3}^{0.7}\langle\cdot\rangle_{\perp} dz$ is the bulk averaging operator. In other words, the heat and salt fluxes can be diagnosed at the ice-ocean interface simply from $\tilde{T}_*$ and $\tekTs$ in the bulk, since we assume steady state. Note that equations \eqref{eq:fluxrec} are valid in the limit $\overline{\tS}(\tz=1)\approx 0$ only (cf. appendix), such that the exact dependence of the interfacial salinity on $Re_{\tau}$, $Pr$ and $Le$ (cf. figure \ref{fig4}) cannot be captured by the reduced model.}

\subsection{Results}\label{sec:tdresults}

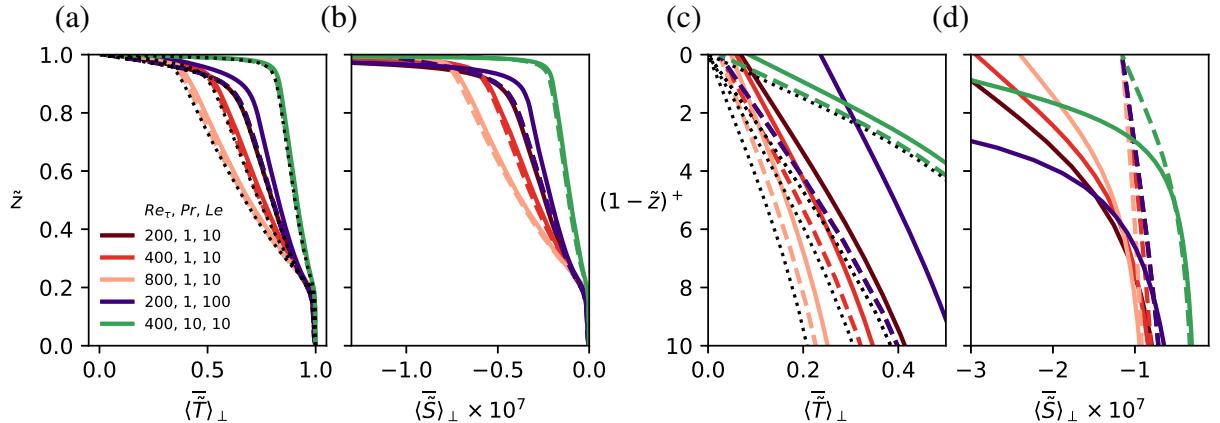
\begin{figure}[t]
  \noindent
  \centering
  \includegraphics[height=5.4cm]{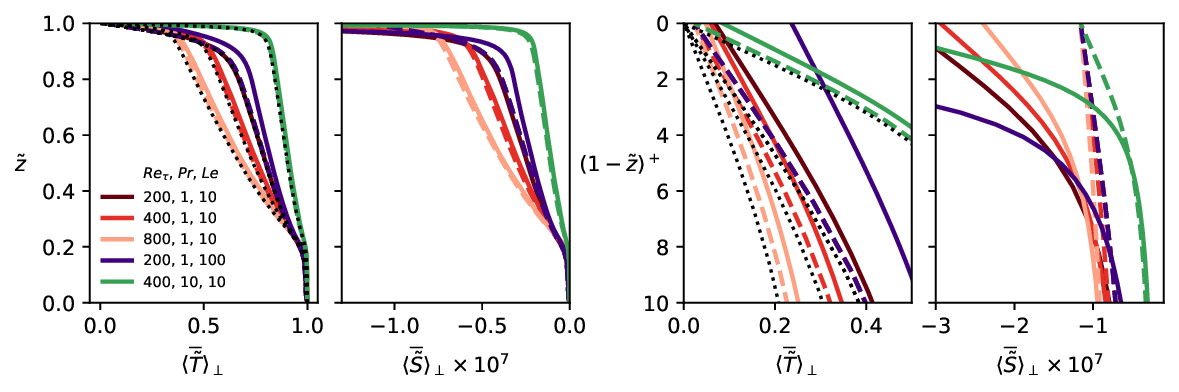}
  \put(-440,155){(a)}  
  \put(-340,155){(b)}  
  \put(-210,155){(c)} 
  \put(-110,155){(d)}  
  \caption{{(a) Mean horizontal temperature and (b) salinity profiles. The thick solid and dashed lines show results from full $T-S$ and reduced $T_*$ model simulations (with $\tT$ and $\tS$ reconstructed from equations \eqref{eq:bulkrec}), respectively, while the dotted black lines in (a) show thermal driving from the reduced $T_*$ model simulations. (c) and (d) are zoomed-in versions of (a) and (b) with the distance from the ice-ocean interface $(1-\tilde{z})^+=Re_{\tau}(1-\tilde{z})$ (in wall units and increasing from top to bottom) replacing the depth coordinate $\tilde{z}$ on the $y$ axis. Note that $(1-\tilde{z})^+=10$ corresponds to $\tilde{z}=0.95$ for $Re_{\tau}=200$.}}\label{fig5}
\end{figure}

Figures \ref{fig5}(a)-(b) show the mean temperature and salinity profiles with depth, respectively, and demonstrate that the reduced $T_*$ model can reproduce some of the key features of the full $T-S$ model. Specifically, the good overlap of the mean temperature and salinity profiles obtained from $T-S$ simulation results (solid lines) with those diagnosed from $T_*$ simulation results (dashed lines) in the bulk validates the MML theory. Figures \ref{fig5}(c)-(d) zoom in on the ice-ocean diffusive sublayers and show that the ${\tT^{diag}}$ and ${\tS^{diag}}$ curves deviate in the boundary layer from the true profiles obtained with the full $T-S$ model. This boundary layer disagreement is expected since the MML theory only applies in the turbulent bulk and illustrates why the MML prediction cannot be used alone to estimate the interfacial heat and salt fluxes. {In addition, the discrepancy increases as $Le$ increases since $Le$ effects are missing in the reduced model (as seen for $Le=100$ from the separation of the solid purple line and dashed purple line, which overlaps with the brown dashed line, in figures \ref{fig5}(a)-(d)).} We note that the thermal driving profiles (shown by the black dotted lines in figures \ref{fig5}(a),(c)) are close to the temperature profiles diagnosed from equation \eqref{eq:bulkrec} in all our cases. This is because $St \gg |\gamma |$, which should always be the case in the oceanographic context. This overlap is observed not only in the bulk but also in the boundary layer when $Pr_*=Pr$. However, we note that it is not a good idea to use $\langle \overline{ \tpz\tTs (\tz=1) }\rangle_{\perp}$ as a proxy for $\langle \overline{ \tpz\tT (\tz=1) }\rangle_{\perp}$. There is no theoretical basis and a more physically-motivated estimate is readily available from equation \eqref{eq:fluxrec}.

\begin{figure}[t]
  \noindent
  \centering
  \includegraphics[height=5.4cm]{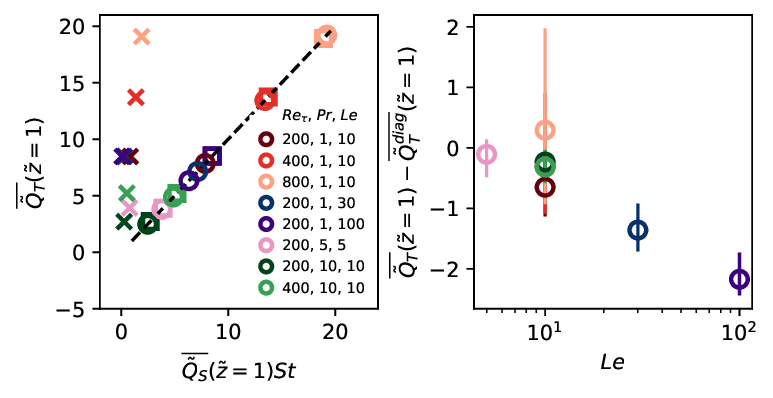}
  \put(-280,155){(a)} 
  \put(-140,155){(b)}  
  \caption{(a) Mean normalized heat flux as a function of the mean normalized salinity flux at the ice-ocean interface obtained directly from full $T-S$ simulation results (unfilled circles) or diagnosed from $T_*$ model results using equations \eqref{eq:fluxrec} and a prescribed $Le$ (unfilled squares). The mean heat and salinity fluxes computed from the gradient of the MML-reconstructed temperature and salinity profiles (cf. \eqref{eq:bulkrec}) are also shown (filled triangles) and unsurprisingly are significantly different from the wall gradients obtained from the full $T-S$ simulations. The dashed line highlights the equation $\tilde{Q}_T=\tilde{Q}_SSt$, which is the melt-induced dilution equation in the limit of small interfacial salt deficit. (b) Difference between the interfacial heat flux estimated from $T-S$ simulation results and $T_*$ simulation results. The vertical bars show the standard deviation for the heat flux based on the time series of the full $T-S$ simulation results (note that the standard deviation measured from the time series of the reduced $T_*$ simulations is typically smaller).}\label{fig6}
\end{figure}

Figure \ref{fig6}(a) shows the mean interfacial fluxes obtained from the full $T-S$ simulation results (unfilled circles) and diagnosed from the $T_*$ model results (unfilled squares) using equations \eqref{eq:fluxrec}. The overall agreement is good. {For $Le=10$, there is a perfect overlap of the circles and squares. For higher $Le$, however, differences can be seen between the exact and diagnosed fluxes. Specifically, the exact fluxes ($\overline{\tilde{Q}_T}$ and $\overline{\tilde{Q}_S}$ at $\tilde{z}=1$) decrease with $Le$ while those diagnosed are, by construction, $Le$ independent (cf. equation \eqref{eq:fluxrec}). This difference is highlighted in figure \ref{fig6}(b). The relative difference is typically 10\% but jumps to 30\% for the high $Le=100$ case. A similar result is obtained for $\overline{\tilde{Q}_S}(\tilde{z}=1)St$ as it is equal to $\overline{\tilde{Q}_T}(\tilde{z}=1)$ when $\overline{\tilde{S}}\ll 1$. We have verified that the relatively coarse resolution of our high $Le$ simulations is not responsible for the flux discrepancy between the two models (see Supplementary Material) and found that the $Le^e=1$ approximation is well satisfied in all cases. Thus, we hypothesize that the decrease of the exact fluxes with $Le$ in the full $T-S$ simulations must be related to the molecular sublayer dynamics. The molecular sublayers are sensitive to the low salt diffusivity and may throttle the vertical heat and salt fluxes, which are quasi linearly correlated at the ice-ocean interface, in the relatively low $Re_{\tau}$ and high $Le$ limit. The lack of a salt diffusivity bottleneck in the $T_*$ model may put a fundamental limit on the range of $Re_{\tau}$ and $Le$ parameters for which it can be used to diagnose the interfacial melt rate. However, the relative error is \textit{only} 30\% for $Re_{\tau}=200$ and $Le=100$, which is relatively close to the ocean value $Le \approx 170$. The error may also decrease with higher (more realistic) $Re_{\tau}$ as the boundary layer bottleneck is thought to vanish when turbulence increases \citep{Iyer2020}.}

{We note that our diagnosis of the interfacial heat and salinity fluxes from $T_*$ model results does not allow to recover the salinity deficit at the ice-ocean interface. Our derivation of $\tilde{Q}_T^{diag}(\tilde{z}=1)$ (cf. equation \eqref{eq:fluxrec}) makes the approximation $1+\overline{\tilde{S}}(\tz=1)\approx 1$, which is valid in the weak thermal driving limit, such that the melt-induced dilution equation based on the diagnosed fluxes yields systematically $\overline{\tilde{S}}(\tz=1)=0$. The absence of a diagnosis for $\overline{\tilde{S}}(\tz=1)$ implies that we cannot diagnose the Nusselt number for the salinity field. However, we can approximate $Nu_T^{diag}=Pr\tilde{Q}_T^{diag}(\tilde{z}=1)\approx Nu_{T_*}$ (assuming $Pr=Pr_*$) since $\tT^{diag}\approx \langle\overline{\tTs}\rangle_{\perp}$ when $St\gg |\gamma|$ (reported in table \ref{tab:numexp}). A diffusive sublayer model would be required to diagnose $\overline{\tilde{S}}(\tz=1)$ but is beyond the scope of the present work.}



\section{Conclusions}

We have shown that the effective Lewis number $Le^e$ is close to unity in direct numerical simulations of temperature- and salt-stratified sub-ice ocean currents (\S\ref{sec2}). {We limited our attention to ambient water masses with spatially uniform and constant-in-time properties, and far-field temperatures close to freezing (weak far-field thermal driving)}. Thus, we have provided supporting evidence for the equal effective diffusivities hypothesis of the MML model in the well-mixed regime. 


Motivated by the observation $Le^e\approx 1$, we have shown how the results from a reduced thermal driving model can be used to diagnose the temperature and salinity profiles in the turbulent bulk and the heat and salt fluxes at the ice-ocean interface (\S\ref{sec3}). {The exact and diagnosed water properties are in good agreement overall, but differ increasingly as the molecular Lewis number $Le$ increases. The high-$Le$ discrepancy is due to the fact that the full $T-S$ simulation results are sensitive to $Le$ while the diagnosed variables are, by construction, $Le$ invariant. The neglect of the interfacial salinity $\overline{\tS(\tz=1)}$ in the derivation of the diagnosed fluxes (equation \eqref{eq:fluxrec}) is responsible for the lack of $Le$ effects in the reconstruction procedure (cf. appendix). Thus, the derivation of a boundary layer model ($Le$ dependent) predicting the interfacial salinity $\overline{\tS(\tz=1)}$ from the reduced model results would improve the diagnoses. We note that the $Le$-induced discrepancy is expected to vanish at high $Re_{\tau}$ (\S\ref{sec3}\ref{sec:tdresults}). }

We listed in table \ref{tab:numexp} key output variables from our simulations. We reiterate that the global quantities we reported, i.e., $Nu_T$, $Nu_S$ and $C_D$, are artificially modified by the small horizontal dimensions of our numerical domains \citep{DeGiovanetti2016}. Thus, simulations in much larger domains will be required to derive scaling laws for $Nu_T$, $Nu_S$ and $C_D$ relevant to the broad (unconfined) ocean. Our aim was not to revisit the parameterization of ice melting in the IOBL in the shear-dominated regime \citep{Holland1999,Malyarenko2020}, but rather to lay the groundwork for reduced models that will be cost-effective for this computationally intensive task.


Future work should examine the minimum far-field thermal driving above which intermittent relaminarization breaks the equal effective diffusivities hypothesis. Simulations with large far-field thermal driving will need to consider large domains as relaminarization is artificially activated in numerically-confined environments \citep{Garcia-Villalba2011}. Previous works suggest that domains with streamwise and spanwise dimensional lengths $6\pi h$ and $2\pi h$, with $h$ the half channel width, or larger, will be necessary to avoid confinement effects \citep{Flores2010,Bernardini2014}. {The effect of an oscillating current would be also worth considering, as tides play an important role in several ice-shelf cavities \citep{Richter2022}. A key question would be whether turbulence survives long enough when tidal currents are low to ensure approximately equal turbulent diffusivities at all times. We remark that some of our approximations may not be valid at high thermal driving. Simulations with relatively high melt rates might need to model deformations of the ice boundary explicitly \citep{Couston2021a}, or have non-zero velocities normal to the interface, and consider a nonlinear equation of state to correctly model density fluctuations in the presencen of large temperature and salinity differences \citep{Hester2021}.}

\acknowledgments This work was granted access to the HPC resources of IDRIS under the allocation 2023-A0140114116 made by GENCI. We gratefully acknowledge support from the PSMN (Pôle Scientifique de Modélisation Numérique) of the ENS de Lyon for the computing resources. 

%
%
\datastatement We will make publicly available (before the paper is published) a Github repository containing all analysis scripts (required to produce the figures) and a Zenodo repository containing both full three-dimensional checkpoint data (at statistical steady state) and post processed data. The software Dedalus is an open-source Python package available from the Github repository https://github.com/DedalusProject/dedalus.

%



\appendix 


\appendixtitle{Derivation of the meltwater mixing line model for a stationary ice-ocean interface}

The MML prediction can be derived from the advection-diffusion equations \eqref{eq:tsmodel3}-\eqref{eq:tsmodel4} for $\tT$ and $\tS$ {assuming a spatially uniform and constant-in-time ambient}. First, we integrate equations \eqref{eq:tsmodel3}-\eqref{eq:tsmodel4} over $(\tx,\ty,\tt)$, assume statistical steady state ($\tpt \equiv 0$) and discard the relaxation terms, which vanish in the bulk, to obtain
\begin{subequations}\label{eq:mmlderA}
\begin{eqnarray}\label{eq:mmlderA1}
&\f{d}{d\tz} \langle \overline{\tw\tT - Pr^{-1}\tpz\tT}\rangle_{\perp} = 0, \\ \label{eq:mmlderA2}
&\f{d}{d\tz} \langle \overline{\tw\tS - Sc^{-1}\tpz\tS}\rangle_{\perp} = 0. 
\end{eqnarray}
\end{subequations}
Then we substitute the definitions for the effective (turbulent) diffusivities \eqref{eq:turbdiff} into \eqref{eq:mmlderA} and integrate from $\tz=1$ (ice-ocean interface) to $\tz<1$ to get 
\begin{subequations}\label{eq:mmlderB}
\begin{eqnarray}\label{eq:mmlderB1}
& \tekT \langle \overline{\tpz\tT}\rangle_{\perp} = Pr^{-1}\langle\overline{\tpz\tT(\tz=1)}\rangle_{\perp}, \\ \label{eq:mmlderB2}
&  \tekS \langle \overline{\tpz\tS}\rangle_{\perp} = Sc^{-1}\langle\overline{\tpz\tS(\tz=1)}\rangle_{\perp}. 
\end{eqnarray}
\end{subequations}
Finally, we take the ratio of equations \eqref{eq:mmlderB1} and \eqref{eq:mmlderB2}, make the assumption of equal effective diffusivities, and use the melt-induced dilution equation \eqref{eq:tsbc} to arrive at the MML prediction, i.e.
\begin{eqnarray}\label{eq:mmlderC}
& \frac{\langle \overline{\tpz\tT}\rangle_{\perp}}{\langle \overline{\tpz\tS}\rangle_{\perp}} = \frac{St}{1+\langle \overline{\tS(\tz=1)} \rangle_{\perp}}. 
\end{eqnarray}

Importantly, the right-hand-side in equation \eqref{eq:mmlderC} can be simplified as $St$ in the limit $\langle \overline{\tS(\tz=1)} \rangle_{\perp} \ll 1$ (small interfacial salt deficit), which is appropriate for all simulation cases discussed in this paper. In this limit, integrating \eqref{eq:mmlderC} from the far-field $\tz=0$ upward yields a simple relation between the mean bulk temperature and salinity, i.e.,
\begin{eqnarray}\label{eq:mmlderD}
& \overline{\tT}-1=\overline{\tS}St. 
\end{eqnarray}
Combining \eqref{eq:mmlderD} with the definition of thermal driving $\tTs=\tT-\gamma \tS$ yields expressions \eqref{eq:bulkrec} for the bulk temperature and salinity diagnosed from $\tTs$.

The correlation coefficient $\mathcal{C}$ derived for a fixed ice-ocean interface (right-hand-side of equation \eqref{eq:mmlderC}) is slightly different from the correlation coefficient derived for a moving interface \citep{Gade1979}. In the latter case, re-doing the analysis in \cite{Gade1979} with dimensionless variables yields $\mathcal{C}\approx 1+St$, which is close to the result obtained for a fixed interface in the limit $St \gg 1$. It is worth noting that our original MML expression \eqref{eq:mmlderC} predicts a correlation coefficient that depends on the interfacial salinity, which itself is a function of the melt rate. This dependence is not physical. It is absent from the MML prediction derived by \cite{Gade1979} and is due to the neglect of meltwater advection across the ice-ocean boundary in our formulation \citep[see a related discussion in section 4 of][]{Jenkins2001}. However, its effect vanishes in the limit of weak thermal driving, which is why all our simulations exhibit the same melt-independent MML slope in figure \ref{fig4}.

%
%
%
%
%
%
%
%
%

\bibliographystyle{ametsocV6}
\bibliography{thermaldriving}

\begin{thebibliography}{50}
\providecommand{\natexlab}[1]{#1}
\providecommand{\url}[1]{\texttt{#1}}
\renewcommand{\UrlFont}{\rmfamily}
\providecommand{\urlprefix}{URL }
\expandafter\ifx\csname urlstyle\endcsname\relax
  \providecommand{\doi}[1]{https://doi.org/\discretionary{}{}{}#1}\else
  \providecommand{\doi}{https://doi.org/\discretionary{}{}{}\begingroup
  \urlstyle{rm}\Url}\fi
\providecommand{\eprint}[2][]{\url{#2}}

\bibitem[{Alc{\'{a}}ntara-{\'{A}}vila and
  Hoyas(2021)Alc{\'{a}}ntara-{\'{A}}vila, and Hoyas}]{Alcantara-Avila2021}
Alc{\'{a}}ntara-{\'{A}}vila, F., and S.~Hoyas, 2021: {Direct numerical
  simulation of thermal channel flow for medium–high Prandtl numbers up to
  Re$\tau$=2000}. \textit{International Journal of Heat and Mass Transfer},
  \textbf{176}, \doi{10.1016/j.ijheatmasstransfer.2021.121412}.

\bibitem[{Begeman et~al.(2022)Begeman, Asay-Davis,, and {Van
  Roekel}}]{Begeman2022}
Begeman, C.~B., X.~Asay-Davis, and L.~{Van Roekel}, 2022: {Ice-shelf ocean
  boundary layer dynamics from large-eddy simulations}. \textit{The
  Cryosphere}, \textbf{16~(1)}, 277--295, \doi{10.5194/tc-16-277-2022}.

\bibitem[{Bernardini et~al.(2014)Bernardini, Pirozzoli,, and
  Orlandi}]{Bernardini2014}
Bernardini, M., S.~Pirozzoli, and P.~Orlandi, 2014: {Velocity statistics in
  turbulent channel flow up to Re$\tau$ =4000}. \textit{Journal of Fluid
  Mechanics}, \textbf{742}, 171--191, \doi{10.1017/jfm.2013.674}.

\bibitem[{Bouffard and Boegman(2013)Bouffard, and Boegman}]{Bouffard2013}
Bouffard, D., and L.~Boegman, 2013: {A diapycnal diffusivity model for
  stratified environmental flows}. \textit{Dynamics of Atmospheres and Oceans},
  \textbf{61-62}, 14--34, \doi{10.1016/j.dynatmoce.2013.02.002},
  \urlprefix\url{http://dx.doi.org/10.1016/j.dynatmoce.2013.02.002}.

\bibitem[{Burns et~al.(2020)Burns, Vasil, Oishi, Lecoanet,, and
  Brown}]{Burns2020}
Burns, K.~J., G.~M. Vasil, J.~S. Oishi, D.~Lecoanet, and B.~P. Brown, 2020:
  {Dedalus: A flexible framework for numerical simulations with spectral
  methods}. \textit{Phys. Rev. Research}, \textbf{2~(2)}, 23\,068,
  \doi{10.1103/PhysRevResearch.2.023068},
  \urlprefix\url{https://link.aps.org/doi/10.1103/PhysRevResearch.2.023068}.

\bibitem[{Cenedese and Straneo(2023)Cenedese, and Straneo}]{Cenedese2023}
Cenedese, C., and F.~Straneo, 2023: {Icebergs Melting}. \textit{Annual Review
  of Fluid Mechanics}, \textbf{55}, 377--402,
  \doi{10.1146/annurev-fluid-032522-100734}.

\bibitem[{Couston et~al.(2021)Couston, Hester, Favier, Taylor, Holland,, and
  Jenkins}]{Couston2021a}
Couston, L.-A., E.~Hester, B.~Favier, J.~R. Taylor, P.~R. Holland, and
  A.~Jenkins, 2021: {Topography generation by melting and freezing in a
  turbulent shear flow}. \textit{Journal of Fluid Mechanics},
  \textbf{911~(A44)}, 1--37, \doi{10.1017/jfm.2020.1064}, \eprint{2004.09879}.

\bibitem[{Davis et~al.(2023)}]{Davis2023}
Davis, P.~E., and Coauthors, 2023: {Suppressed basal melting in the eastern
  Thwaites Glacier grounding zone}. \textit{Nature}, \textbf{614~(7948)},
  479--485, \doi{10.1038/s41586-022-05586-0}.

\bibitem[{{De Giovanetti} et~al.(2016){De Giovanetti}, Hwang,, and
  Choi}]{DeGiovanetti2016}
{De Giovanetti}, M., Y.~Hwang, and H.~Choi, 2016: {Skin-friction generation by
  attached eddies in turbulent channel flow}. \textit{Journal of Fluid
  Mechanics}, \textbf{808}, 511--538, \doi{10.1017/jfm.2016.665}.

\bibitem[{Dinniman et~al.(2016)Dinniman, Asay-Davis, Galton-Fenzi, Holland,
  Jenkins,, and Timmermann}]{Dinniman2016}
Dinniman, M.~S., X.~S. Asay-Davis, B.~K. Galton-Fenzi, P.~R. Holland,
  A.~Jenkins, and R.~Timmermann, 2016: {Modeling ice shelf/ocean interaction in
  Antarctica: A review}. \textit{Oceanography}, \textbf{29~(4)}, 144--153,
  \doi{10.5670/oceanog.2016.106}.

\bibitem[{Falor et~al.(2023)Falor, Gayen, Sengupta,, and Ivey}]{Falor2023}
Falor, D., B.~Gayen, D.~Sengupta, and G.~N. Ivey, 2023: {Evaporation induced
  convection enhances mixing in the upper ocean}. \textit{Frontiers in Marine
  Science}, \textbf{10~(May)}, 1--9, \doi{10.3389/fmars.2023.1176226}.

\bibitem[{Flores and Jim{\'{e}}nez(2010)Flores, and Jim{\'{e}}nez}]{Flores2010}
Flores, O., and J.~Jim{\'{e}}nez, 2010: {Hierarchy of minimal flow units in the
  logarithmic layer}. \textit{Physics of Fluids}, \textbf{22~(7)}, 1--4,
  \doi{10.1063/1.3464157}.

\bibitem[{Gade(1979)}]{Gade1979}
Gade, H.~G., 1979: {Melting of Ice in Sea Water: A Primitive Model with
  Application to the Antarctic Ice Shelf and Icebergs}. \textit{Journal of
  Physical Oceanography}, \textbf{9~(1)}, 189--198,
  \urlprefix\url{https://doi.org/10.1175/1520-0485(1979)009{\%}3C0189:MOIISW{\%}3E2.0.CO;2},
  \eprint{arXiv:1011.1669v3}.

\bibitem[{Garc{\'{i}}a-Villalba and del {\'{A}}lamo(2011)Garc{\'{i}}a-Villalba,
  and del {\'{A}}lamo}]{Garcia-Villalba2011}
Garc{\'{i}}a-Villalba, M., and J.~C. del {\'{A}}lamo, 2011: {Turbulence
  modification by stable stratification in channel flow}. \textit{Physics of
  Fluids}, \textbf{23~(4)}, \doi{10.1063/1.3560359}.

\bibitem[{Gregg et~al.(2018)Gregg, D'Asaro, Riley,, and Kunze}]{Gregg2018}
Gregg, M.~C., E.~A. D'Asaro, J.~J. Riley, and E.~Kunze, 2018: {Mixing
  efficiency in the ocean}. \textit{Annual Review of Marine Science},
  \textbf{10}, 443--473, \doi{10.1146/annurev-marine-121916-063643}.

\bibitem[{Hester et~al.(2021)Hester, McConnochie, Cenedese, Couston,, and
  Vasil}]{Hester2021}
Hester, E.~W., C.~D. McConnochie, C.~Cenedese, L.-A. Couston, and G.~Vasil,
  2021: {Aspect ratio affects iceberg melting}. \textit{Physical Review
  Fluids}, \textbf{6~(2)}, 23\,802, \doi{10.1103/PhysRevFluids.6.023802},
  \urlprefix\url{https://doi.org/10.1103/PhysRevFluids.6.023802},
  \eprint{2009.10281}.

\bibitem[{Hewitt(2020)}]{Hewitt2020}
Hewitt, I.~J., 2020: {Subglacial Plumes}. \textit{Annual Review of Fluid
  Mechanics}, \textbf{52}, 145--169, \doi{10.1146/annurev-fluid-010719-060252}.

\bibitem[{Holland and Jenkins(1999)Holland, and Jenkins}]{Holland1999}
Holland, D.~M., and A.~Jenkins, 1999: {Modeling thermodynamic ice-ocean
  interactions at the base of an ice shelf}. \textit{Journal of Physical
  Oceanography}, \textbf{29}, 1787--1800,
  \doi{10.1175/1520-0485(1999)029<1787:mtioia>2.0.co;2},
  \urlprefix\url{Modeling thermodynamic ice-ocean interactions at the base of
  an ice shelf}.

\bibitem[{Howland et~al.(2023)Howland, Verzicco,, and Lohse}]{Howland2023}
Howland, C.~J., R.~Verzicco, and D.~Lohse, 2023: {Double-diffusive transport in
  multicomponent vertical convection}. \textit{Physical Review Fluids},
  \textbf{8~(1)}, 1--20, \doi{10.1103/PhysRevFluids.8.013501}.

\bibitem[{Hwang(2013)}]{Hwang2013}
Hwang, Y., 2013: {Near-wall turbulent fluctuations in the absence of wide outer
  motions}. \textit{Journal of Fluid Mechanics}, \textbf{723}, 264--288,
  \doi{10.1017/jfm.2013.133}.

\bibitem[{Iyer et~al.(2020)Iyer, Scheel, Schumacher,, and
  Sreenivasan}]{Iyer2020}
Iyer, K.~P., J.~D. Scheel, J.~Schumacher, and K.~R. Sreenivasan, 2020:
  {Classical 1/3 scaling of convection holds up to Ra = 1015}.
  \textit{Proceedings of the National Academy of Sciences of the United States
  of America}, \textbf{117~(14)}, 7594--7598, \doi{10.1073/pnas.1922794117}.

\bibitem[{Jackson and Rehmann(2003)Jackson, and Rehmann}]{Jackson2003}
Jackson, P.~R., and C.~R. Rehmann, 2003: {Laboratory Measurements of
  Differential Diffusion in a Diffusively Stable, Turbulent Flow}.
  \textit{Journal of Physical Oceanography}, \textbf{33~(8)}, 1592--1603,
  \doi{10.1175/1520-0485(2003)033<1592:lmoddi>2.0.co;2}.

\bibitem[{Jenkins(1999)}]{Jenkins1999}
Jenkins, A., 1999: {The impact of melting ice on ocean waters}. \textit{Journal
  of Physical Oceanography}, \textbf{29~(9)}, 2370--2381,
  \doi{10.1175/1520-0485(1999)029<2370:TIOMIO>2.0.CO;2}.

\bibitem[{Jenkins(2011)}]{Jenkins2011}
Jenkins, A., 2011: {Convection-driven melting near the grounding lines of ice
  shelves and tidewater glaciers}. \textit{Journal of Physical Oceanography},
  \textbf{41~(12)}, 2279--2294, \doi{10.1175/JPO-D-11-03.1}.

\bibitem[{Jenkins(2016)}]{Jenkins2016}
Jenkins, A., 2016: {A simple model of the ice shelf-ocean boundary layer and
  current}. \textit{Journal of Physical Oceanography}, \textbf{46~(6)},
  1785--1803, \doi{10.1175/JPO-D-15-0194.1}.

\bibitem[{Jenkins(2021)}]{Jenkins2021}
Jenkins, A., 2021: {Shear, Stability and Mixing within the Ice-Shelf-Ocean
  Boundary Current}. \textit{Journal of Physical Oceanography}, 2129--2148,
  \doi{10.1175/jpo-d-20-0096.1}.

\bibitem[{Jenkins et~al.(2001)Jenkins, Hellmer,, and Holland}]{Jenkins2001}
Jenkins, A., H.~H. Hellmer, and D.~M. Holland, 2001: {The role of meltwater
  advection in the formulation of conservative boundary conditions at an
  Ice-Ocean interface}. \textit{Journal of Physical Oceanography},
  \textbf{31~(1)}, 285--296,
  \doi{10.1175/1520-0485(2001)031<0285:TROMAI>2.0.CO;2}.

\bibitem[{Jenkins et~al.(2010)Jenkins, Nicholls,, and Corr}]{Jenkins2010}
Jenkins, A., K.~W. Nicholls, and H.~F. Corr, 2010: {Observation and
  parameterization of ablation at the base of Ronne Ice Ahelf, Antarctica}.
  \textit{Journal of Physical Oceanography}, \textbf{40~(10)}, 2298--2312,
  \doi{10.1175/2010JPO4317.1}.

\bibitem[{Jim{\'{e}}nez and Moin(1991)Jim{\'{e}}nez, and Moin}]{Jimenez1991}
Jim{\'{e}}nez, J., and P.~Moin, 1991: {The minimal flow unit in near-wall
  turbulence}. \textit{Journal of Fluid Mechanics}, \textbf{225}, 213--240,
  \doi{10.1017/S0022112091002033}.

\bibitem[{Kimura et~al.(2016)Kimura, Jenkins, Dutrieux, Forryan, Garabato,, and
  Firing}]{Kimura2016}
Kimura, S., A.~Jenkins, P.~Dutrieux, A.~Forryan, A.~C.~N. Garabato, and
  Y.~Firing, 2016: {Ocean mixing beneath Pine Island Glacier ice shelf, West
  Antarctica}. \textit{Journal of Geophysical Research: Oceans},
  \textbf{121~(5)}, 3010--3028, \doi{10.1002/2016JC012149.Received}.

\bibitem[{Kimura et~al.(2015)Kimura, Nicholls,, and Venables}]{Kimura2015}
Kimura, S., K.~W. Nicholls, and E.~Venables, 2015: {Estimation of ice shelf
  melt rate in the presence of a thermohaline staircase}. \textit{Journal of
  Physical Oceanography}, \textbf{45~(1)}, 133--148,
  \doi{10.1175/JPO-D-14-0106.1}.

\bibitem[{Lozano-Dur{\'{a}}n and Jim{\'{e}}nez(2014)Lozano-Dur{\'{a}}n, and
  Jim{\'{e}}nez}]{LozanoDuran2014}
Lozano-Dur{\'{a}}n, A., and J.~Jim{\'{e}}nez, 2014: {Effect of the
  computational domain on direct simulations of turbulent channels up to
  Re$\tau$ = 4200}. \textit{Physics of Fluids}, \textbf{26~(1)},
  \doi{10.1063/1.4862918}.

\bibitem[{Ma and Peltier(2022)Ma, and Peltier}]{Ma2022b}
Ma, Y., and W.~R. Peltier, 2022: {Diapycnal diffusivities in Kelvin-Helmholtz
  engendered turbulent mixing: The diffusive-convection regime in the Arctic
  Ocean}. \textit{Journal of Fluid Mechanics}, \textbf{946}, 1--35,
  \doi{10.1017/jfm.2022.590}.

\bibitem[{Malyarenko et~al.(2020)Malyarenko, Wells, Langhorne, Robinson,
  Williams,, and Nicholls}]{Malyarenko2020}
Malyarenko, A., A.~J. Wells, P.~J. Langhorne, N.~J. Robinson, M.~J.~M.
  Williams, and K.~W. Nicholls, 2020: {A synthesis of thermodynamic ablation at
  ice–ocean interfaces from theory, observations and models}. \textit{Ocean
  Modelling}, \textbf{154}, 101\,692,
  \doi{https://doi.org/10.1016/j.ocemod.2020.101692},
  \urlprefix\url{https://doi.org/10.1016/j.ocemod.2020.101692}.

\bibitem[{Martin and Rehmann(2006)Martin, and Rehmann}]{Martin2006}
Martin, J.~E., and C.~R. Rehmann, 2006: {Layering in a flow with diffusively
  stable temperature and salinity stratification}. \textit{Journal of Physical
  Oceanography}, \textbf{36~(7)}, 1457--1470, \doi{10.1175/JPO2920.1}.

\bibitem[{McDougall et~al.(2014)McDougall, Barker, Feistel,, and
  Galton-Fenzi}]{McDougall2014}
McDougall, T.~J., P.~M. Barker, R.~Feistel, and B.~K. Galton-Fenzi, 2014:
  {Melting of ice and sea ice into seawater and frazil ice formation}.
  \textit{Journal of Physical Oceanography}, \textbf{44~(7)}, 1751--1775,
  \doi{10.1175/JPO-D-13-0253.1}.

\bibitem[{Middleton et~al.(2021)Middleton, Vreugdenhil, Holland,, and
  Taylor}]{Middleton2021}
Middleton, L., C.~A. Vreugdenhil, P.~R. Holland, and J.~R. Taylor, 2021:
  {Numerical Simulations of Melt-Driven Double-Diffusive Fluxes in a Turbulent
  Boundary Layer beneath an Ice Shelf}. \textit{Journal of Physical
  Oceanography}, \textbf{51~(2)}, 403--418, \doi{10.1175/jpo-d-20-0114.1}.

\bibitem[{Mondal et~al.(2019)Mondal, Gayen, Griffiths,, and Kerr}]{Mondal2019}
Mondal, M., B.~Gayen, R.~W. Griffiths, and R.~C. Kerr, 2019: {Ablation of
  sloping ice faces into polar seawater}. \textit{Journal of Fluid Mechanics},
  \textbf{863}, 545--571, \doi{10.1017/jfm.2018.970}.

\bibitem[{Patmore et~al.(2023)Patmore, Holland, Vreugdenhil,, and
  Jenkins}]{Patmore2022}
Patmore, R.~D., P.~R. Holland, C.~A. Vreugdenhil, and A.~Jenkins, 2023:
  {Turbulence in the ice shelf-ocean boundary current and its sensitivity to
  model resolution}. \textit{Journal of Physical Oceanography}, \textbf{53},
  613 -- 633.

\bibitem[{Pope(2000)}]{Pope2000}
Pope, S.~B., 2000: \textit{{Turbulent Flows}}. Cambridge University Press.

\bibitem[{Reynolds(1975)}]{Reynolds1975}
Reynolds, A.~J., 1975: {The prediction of turbulent Prandtl and Schmidt
  numbers}. \textit{International Journal of Heat and Mass Transfer},
  \textbf{18~(9)}, 1055--1069, \doi{10.1016/0017-9310(75)90223-9}.

\bibitem[{Richter et~al.(2022)Richter, Gwyther, King,, and
  Galton-Fenzi}]{Richter2022}
Richter, O., D.~E. Gwyther, M.~A. King, and B.~K. Galton-Fenzi, 2022: {The
  impact of tides on Antarctic ice shelf melting}. \textit{Cryosphere},
  \textbf{16~(4)}, 1409--1429, \doi{10.5194/tc-16-1409-2022}.

\bibitem[{Rosevear et~al.(2022{\natexlab{a}})Rosevear, Galton-Fenzi,, and
  Stevens}]{Rosevear2022a}
Rosevear, M., B.~Galton-Fenzi, and C.~Stevens, 2022{\natexlab{a}}: {Evaluation
  of basal melting parameterisations using in situ ocean and melting
  observations from the Amery Ice Shelf, East Antarctica}. \textit{Ocean
  Science}, \textbf{18}, 1109--1130,
  \urlprefix\url{https://doi.org/10.5194/os-2021-111}.

\bibitem[{Rosevear et~al.(2021)Rosevear, Gayen,, and
  Galton-Fenzi}]{Rosevear2021}
Rosevear, M.~G., B.~Gayen, and B.~K. Galton-Fenzi, 2021: {The role of
  double-diffusive convection in basal melting of Antarctic ice shelves}.
  \textit{Proceedings of the National Academy of Sciences}, \textbf{118~(6)},
  \doi{10.1073/pnas.2007541118},
  \urlprefix\url{https://www.pnas.org/content/118/6/e2007541118}.

\bibitem[{Rosevear et~al.(2022{\natexlab{b}})Rosevear, Gayen,, and
  Galton-Fenzi}]{Rosevear2022b}
Rosevear, M.~G., B.~Gayen, and B.~K. Galton-Fenzi, 2022{\natexlab{b}}: {Regimes
  and transitions in the basal melting of Antarctic ice shelves}.
  \textit{Journal of Physical Oceanography}, \textbf{52}, 2589 -- 2608,
  \doi{10.1175/jpo-d-21-0317.1}.

\bibitem[{Smyth et~al.(2005)Smyth, Nash,, and Moum}]{Smyth2005}
Smyth, W.~D., J.~D. Nash, and J.~N. Moum, 2005: {Differential diffusion in
  breaking Kelvin-Helmholtz billows}. \textit{Journal of Physical
  Oceanography}, \textbf{35~(6)}, 1004--1022, \doi{10.1175/JPO2739.1}.

\bibitem[{Stevens et~al.(2020)Stevens, Hulbe, Brewer, Stewart, Robinson,
  Ohneiser,, and Jendersie}]{Stevens2020}
Stevens, C., C.~Hulbe, M.~Brewer, C.~Stewart, N.~Robinson, C.~Ohneiser, and
  S.~Jendersie, 2020: {Ocean mixing and heat transport processes observed under
  the Ross Ice Shelf control its basal melting}. \textit{Proceedings of the
  National Academy of Sciences of the United States of America},
  \textbf{117~(29)}, 16\,799--16\,804, \doi{10.1073/pnas.1910760117}.

\bibitem[{Vreugdenhil and Taylor(2019)Vreugdenhil, and
  Taylor}]{Vreugdenhil2019}
Vreugdenhil, C.~A., and J.~R. Taylor, 2019: {Stratification effects in the
  turbulent boundary layer beneath a melting ice shelf: Insights from resolved
  large-eddy simulations}. \textit{Journal of Physical Oceanography},
  \textbf{49~(7)}, 1905--1925, \doi{10.1175/JPO-D-18-0252.1}.

\bibitem[{Vreugdenhil et~al.(2022)Vreugdenhil, Taylor, Davis, Nicholls,
  Holland,, and Jenkins}]{Vreugdenhil2022}
Vreugdenhil, C.~A., J.~R. Taylor, P.~E.~D. Davis, K.~W. Nicholls, P.~R.
  Holland, and A.~Jenkins, 2022: {The ocean boundary layer beneath Larsen C Ice
  Shelf: insights from large-eddy simulations with a near-wall model}.
  \textit{Journal of Physical Oceanography}, \doi{10.1175/jpo-d-21-0166.1}.

\bibitem[{Zhou et~al.(2017)Zhou, Taylor,, and Caulfield}]{Zhou2017}
Zhou, Q., J.~R. Taylor, and C.~P. Caulfield, 2017: {Self-similar mixing in
  stratified plane Couette flow for varying Prandtl number}. \textit{Journal of
  Fluid Mechanics}, \textbf{820}, 86--120, \doi{10.1017/jfm.2017.200}.

\end{thebibliography}

\newpage

\begin{center}
{\Large Supplementary Material} \\ 
{\large -------------------------------- } \\
{\large \textbf{Turbulent ice-ocean boundary layers in the well-mixed regime: \\ insights from direct numerical simulations} }\\ 
{\large -------------------------------- } \\
{\large Louis-Alexandre Couston$^a$} \\
{\large $^a$\textit{ENSL, UCBL, CNRS, Laboratoire de physique, F-69342 Lyon, France}} \\
{\large -------------------------------- } \\
\end{center} 

\textit{In this Supplementary Material we provide details about the numerical resolution of the simulations and show that the key results of the main text are not sensitive to resolution changes.} \\

We compare the grid spacing of our simulations with the turbulent Kolmogorov and Batchelor length scales \citep{Pope2000}, defined as
\begin{eqnarray}\label{eq:kolmo}
\tilde{\eta}_K = \frac{1}{\langle\overline{\tilde{\epsilon}}\rangle_{\perp,\tilde{z}>0.5}^{1/4}}, \quad \tilde{\eta}_{B} = \frac{\tilde{\eta}_K}{Sc^{1/2}},
\end{eqnarray}
The later is based on $Sc$ rather than $Pr$, such that it provides an estimate of the smallest (salinity) features in the flow \citep{Falor2023}. Here $\langle\overline{\tilde{\epsilon}}\rangle_{\perp,\tilde{z}>0.5}$ is the (pseudo) turbulent kinetic energy dissipation rate averaged over the upper half of the domain, i.e.
\begin{eqnarray}\label{eq:tked}
\langle\overline{\tilde{\epsilon}}\rangle_{\perp,\tilde{z}>0.5} = \langle\overline{ \frac{\partial \tilde{\bm{u}}^{\prime}}{\partial \tilde{x}}\cdot\frac{\partial \tilde{\bm{u}}^{\prime}}{\partial \tilde{x}} + \frac{\partial \tilde{\bm{u}}^{\prime}}{\partial \tilde{y}}\cdot\frac{\partial \tilde{\bm{u}}^{\prime}}{\partial \tilde{y}} + \frac{\partial \tilde{\bm{u}}^{\prime}}{\partial \tilde{z}}\cdot\frac{\partial \tilde{\bm{u}}^{\prime}}{\partial \tilde{z}} }\rangle_{\perp,\tilde{z}>0.5},
\end{eqnarray} 
where the prime denotes fluctuations relative to the horizontal mean. The relative difference between $\langle\overline{\tilde{\epsilon}}\rangle_{\perp,\tilde{z}>0.5}$ and the full (averaged) turbulent kinetic energy dissipation rate is much less than $10^{-2}$, hence negligible. Note that $\tilde{\eta}_K$ and $\tilde{\eta}_B$ vary only by about 10\% and always less than 30\% when replacing the volume averaging operator in \eqref{eq:tked} with another operator, such as the horizontal average evaluated at the depth of maximum turbulent kinetic energy. 

To further asses whether our simulations are adequately resolved we count the number of grid points lying within the momentum, thermal and haline boundary layers. We define the lower edge of each boundary layer as the position where the horizontally-averaged diffusive and convective fluxes become equal. Mathematically, we obtain the thickness of the momentum, thermal and haline boundary layers, which we denote by $\tilde{\delta}_u$, $\tilde{\delta}_T$ and $\tilde{\delta}_S$, respectively, by ensuring that they satisfy
\begin{subequations}\label{eq:blthick}
\begin{eqnarray}
\langle\overline{\tilde{w}\tilde{u}}\rangle_{\perp} = & -\langle\overline{\partial_{\tilde{z}}\tilde{u}}\rangle_{\perp}, \quad & \text{at} \quad \tilde{z}=1-\tilde{\delta}_u, \\
\langle\overline{\tilde{w}\tilde{T}}\rangle_{\perp} = & -Pr^{-1}\langle\overline{\partial_{\tilde{z}}\tilde{T}}\rangle_{\perp}, \quad & \text{at} \quad \tilde{z}=1-\tilde{\delta}_T, \\
\langle\overline{\tilde{w}\tilde{S}}\rangle_{\perp} = & -Sc^{-1}\langle\overline{\partial_{\tilde{z}}\tilde{S}}\rangle_{\perp}, \quad & \text{at} \quad \tilde{z}=1-\tilde{\delta}_S.
\end{eqnarray}
\end{subequations}
As in the main text we denote by subscript $+$ variables pre-multiplied by $Re_{\tau}$ (i.e. expressed in terms of wall units).
%

\begin{figure}[t]
\centering
\includegraphics[height=6.8cm]{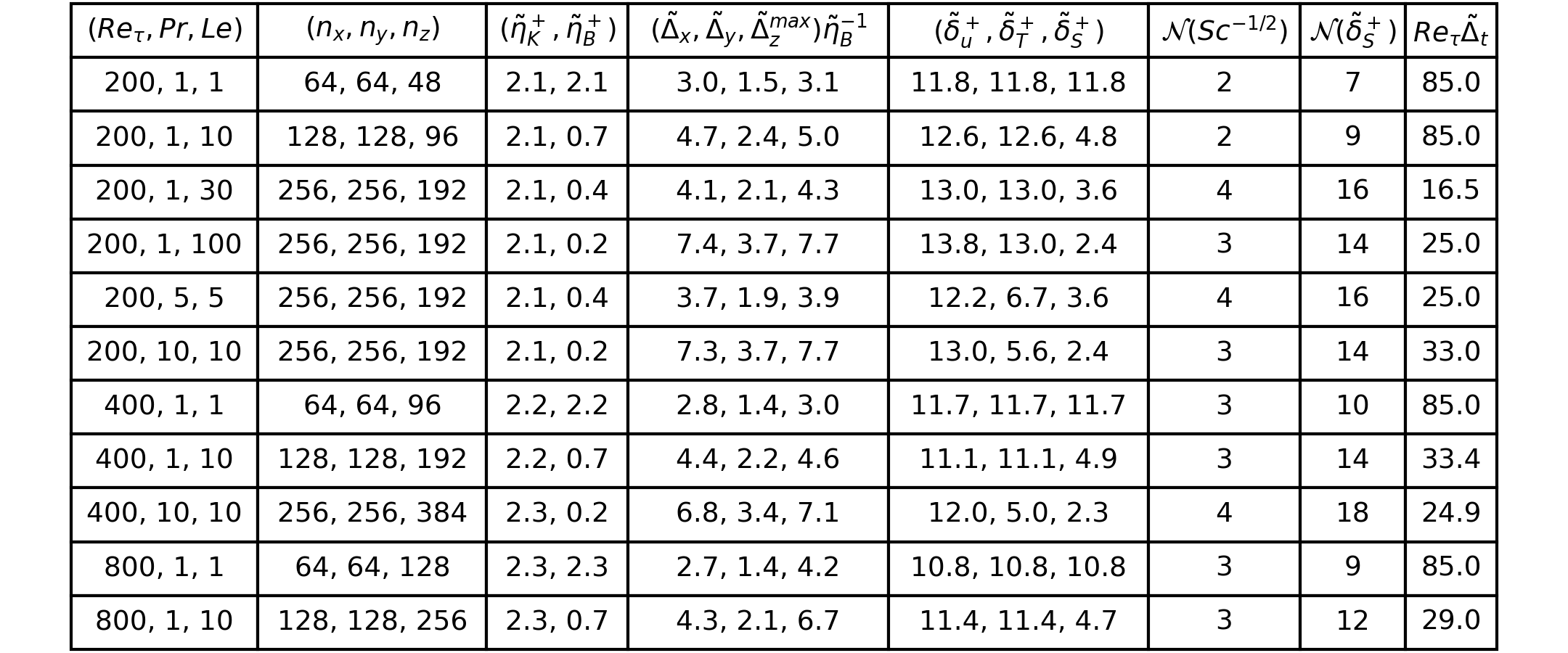}
\caption{Table comparing the grid spacing against the smallest length scales of the flow for all Full $T-S$ simulations discussed in the main text. Note that the typical time step $Re_{\tau}d\tilde{t}$ varies between $5\times 10^{-5}$ and $6\times 10^{-4}$ across all simulations.}
\label{sifig1}
\end{figure}

Figure/Table \ref{sifig1} shows for each Full $T-S$ simulation discussed in the main text the grid resolution (without the 3/2 dealiasing factor), the friction Kolmogorov and Batchelor length scales, the grid spacing normalized by the Batchelor length scale, the boundary layer thicknesses, the number of grid points $\mathcal{N}(Sc^{-1/2})$ and $\mathcal{N}(\tilde{\delta}_S^+)$ satisfying $1-\tilde{z}^+<Sc^{-1/2}$ and $1-\tilde{z}^+<\tilde{\delta}_S^+$, respectively, and the temporal range of time-averaged statistics (in friction time units). 

The Chebyshev collocation points cluster sufficiently close to boundaries for the first grid point to always lie within the distance $1/(Re_{\tau}Sc^{1/2})$ of the ice-ocean interface and for about 10 or more points to lie within the haline molecular sublayer (6th and 7th columns). However, the fourth column clearly shows that our simulations only coarsely resolve the Batchelor scale in the bulk. Most notably, the simulations with the highest Lewis number have grid spacings in $\tilde{x}$ and $\tilde{z}$ that are 7.7 times the Batchelor length scale, exceeding by a factor $\sim 2$ the generally-accepted criterion for resolution requirement \citep{Falor2023}. We note that coarsely resolving the Batchelor length scale is not uncommon \citep{Middleton2021}. In fact little kinetic energy is expected at and below the Kolmogorov length scale such that numerical errors can be expected to remain at small scales.

\begin{figure}[t]
\centering
\includegraphics[height=5cm]{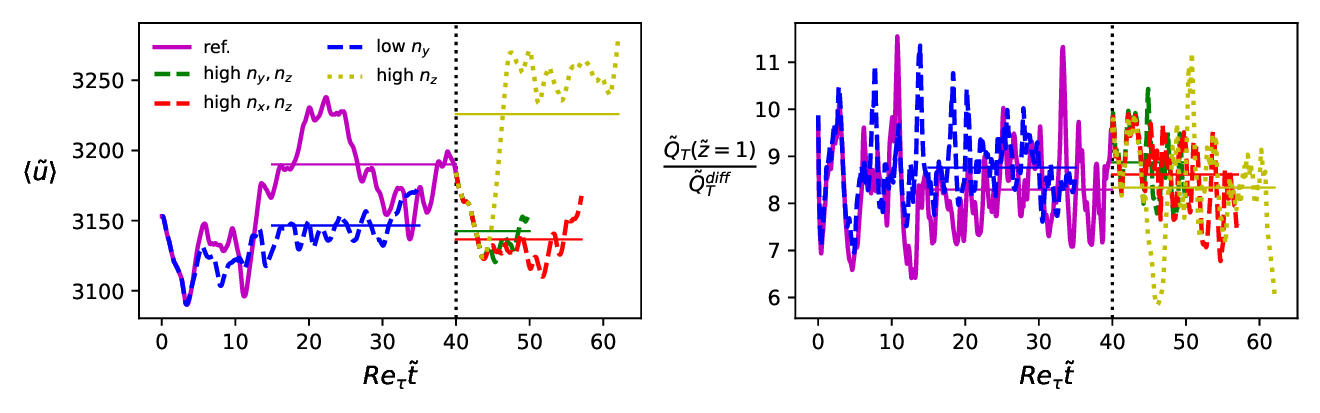}
\put(-430,140){\large{(a)}}
\put(-210,140){\large{(b)}}
\vspace{-0.15in}
\caption{Time series of (a) the mean streamwise flow and (b) the normalized heat flux for the same set of physical parameters, i.e., $Re_{\tau}=200$, $Pr=1$, $Le=100$ but five different spectral resolutions (see table \ref{sifig3}). The simulations with higher resolution than the reference simulation (solid magenta) start from the reference simulation's last checkpoint at time $Re_{\tau}\tilde{t}=40$ (vertical dotted line). The horizontal lines show the mean value over the temporal window used for time averaging.}
\label{sifig2}
\end{figure}

\begin{figure}[t]
\centering
\includegraphics[height=3.6cm]{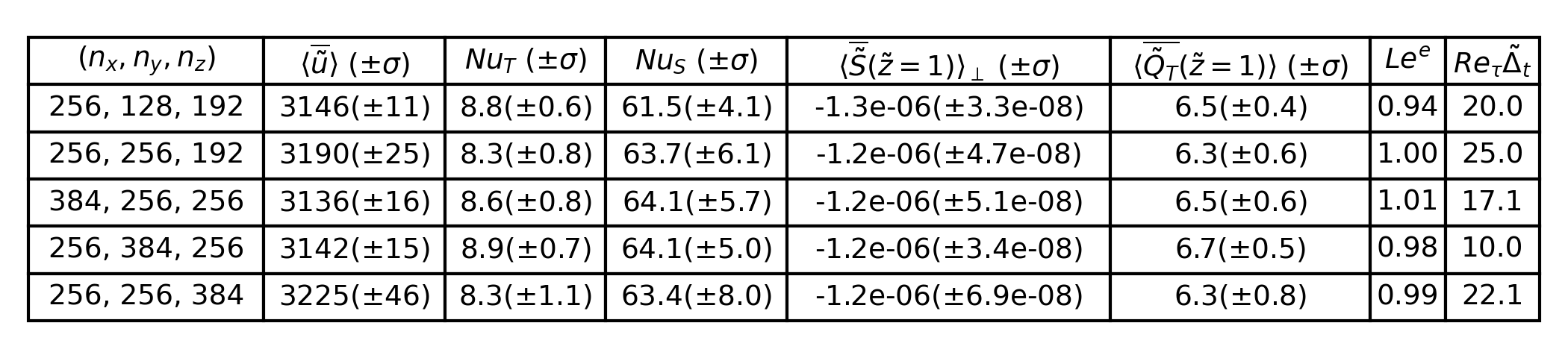}
\vspace{-0.15in}
\caption{Table showing simulation results with $Re_{\tau}=200$, $Pr=1$, $Le=100$ for a range of spectral resolutions. The results of the reference simulation, discussed in the main text, are on the second line ($n_x,n_y,n_z$=256,256,192). $\sigma$ denotes the standard deviation in time.}
\label{sifig3}
\end{figure}

While our simulations do not visibly exhibit non-physical behaviours despite the coarse resolution, except numerical ringing occasionally, we investigate the impact of changing the resolution on a handful of statistics for one of the simulations with the coarsest (relative) resolution, i.e. $Re_{\tau}=200$, $Pr=1$, $Le=100$. Figure \ref{sifig2} shows the mean flow rate and normalized heat flux at the ice-ocean interface over time for five different resolutions (cf. figure/table \ref{sifig3}). The curves for the lower resolution and reference simulations overlap almost perfectly from  $Re_{\tau}\tilde{t}=0$ to $Re_{\tau}\tilde{t}\approx$5, while those at higher resolution (restarted from the reference simulation at time $Re_{\tau}\tilde{t}=40$) overlap from $Re_{\tau}\tilde{t}=40$ to $Re_{\tau}\tilde{t}\approx$45. At later times the results diverge but the mean value remains close to or within the standard deviation (in time) of the reference simulation's result. The relatively negligible impact of resolution on first order statistics is generalized in figure/table \ref{sifig3}. The mean value of all key statistics is close to or within the standard deviation of the reference simulation's result in all cases. Most importantly, the turbulent Lewis number is always close to 1.  

We conclude that the key results of the main text, including $Le^e\approx 1$, are not sensitive to resolution changes. Some of the first order statistics may vary by up to one standard deviation (typically 10\% or less). However, similar changes may be obtained with longer time averaging windows as most of our simulations are relatively short. Having first order statistics converged to a sufficiently high level of accuracy to derive scaling laws is beyond the scope of the manuscript and reserved for future work.

\end{document}